\documentclass[10pt,journal,compsoc]{IEEEtran}

\ifCLASSOPTIONcompsoc
\usepackage[nocompress]{cite}
\else
\usepackage{cite}
\fi
\usepackage{soul,color}
\definecolor{redd}{rgb}{0.858, 0.188, 0.478}
\usepackage{tcolorbox}
\tcbuselibrary{listings,breakable}
\usepackage{algorithm,algorithmic,amsmath, amssymb}
\usepackage{pifont}
\usepackage{microtype}
\usepackage{times}
\usepackage{soul}
\usepackage{url}
\usepackage[utf8]{inputenc}
\usepackage{graphicx}
\usepackage{amsmath}
\usepackage{amsthm}
\usepackage{booktabs}
\usepackage{algorithm}
\usepackage{algorithmic}
\usepackage{xspace}
\newcommand{\tool}{GraphSearchNet\xspace}
\usepackage{color}
\usepackage{amsfonts}       
\usepackage[switch]{lineno}
\usepackage{multirow}
\usepackage{enumitem}
\usepackage{caption}
\usepackage{subcaption}
\usepackage{listings}
\usepackage{float}
\usepackage[hidelinks]{hyperref}
\definecolor{codegreen}{rgb}{0,0.6,0}
\definecolor{codegray}{rgb}{0.5,0.5,0.5}
\definecolor{codepurple}{rgb}{0.58,0,0.82}
\definecolor{backcolour}{rgb}{0.95,0.95,0.92}

\newcommand{\revised}[1][\textcolor{black}]{#1}
\makeatletter
\DeclareRobustCommand\onedot{\futurelet\@let@token\@onedot}
\def\@onedot{\ifx\@let@token.\else.\null\fi\xspace}
\def\eg{\emph{e.g}\onedot} 
\def\ie{\emph{i.e}\onedot}

\makeatother
\usepackage{makecell}
\usepackage{hyperref}
\usepackage{balance}
\begin{document}
\title{GraphSearchNet: Enhancing GNNs via Capturing Global Dependencies for \\ Semantic Code Search}

\author{Shangqing Liu, 
        Xiaofei Xie,
        Jingkai Siow,
        Lei Ma,
        Guozhu Meng
        and Yang Liu 
\IEEEcompsocitemizethanks{\IEEEcompsocthanksitem Shangqing Liu, Jingkai Siow, Yang Liu are with Nanyang Technological University, Singapore. \protect
E-mail: shangqin001@e.ntu.edu.sg, jingkai001@e.ntu.edu.sg, yangliu@ntu.edu.sg
\IEEEcompsocthanksitem Xiaofei Xie is with Singapore Management University, Singapore. E-mail: xiaofei.xfxie@gmail.com
\IEEEcompsocthanksitem Lei Ma is with University of Alberta, Canada. E-mail: ma.lei@acm.org
\IEEEcompsocthanksitem Guozhu Meng is with SKLOIS, Institute of Information Engineering, Chinese Academy of Sciences, China. E-mail: mengguozhu@iie.ac.cn
\IEEEcompsocthanksitem Xiaofei Xie is the corresponding author.
}
\thanks{Manuscript received February 5, 2022}}

\markboth{Journal of \LaTeX\ Class Files,~Vol.0, No.0, February~2022}%
{Liu \MakeLowercase{{et al.}}: \tool: GraphSearchNet: Enhancing GNNs via Capturing Global Dependencies for  Semantic Code Search}

\IEEEtitleabstractindextext{%
\begin{abstract}
Code search aims to retrieve accurate code snippets based on a natural language query to improve software productivity and quality.
With the massive amount of available programs such as (on GitHub or Stack Overflow), identifying and localizing the precise code is critical for the software developers. In addition, Deep learning has recently been widely applied to different code-related scenarios, \eg, vulnerability detection, source code summarization. However, automated deep code search is still challenging \revised{since it requires a high-level semantic mapping between code and natural language queries}. Most existing deep learning-based approaches for code search rely on the sequential text \ie, feeding the program and the query as a flat sequence of tokens to learn the program semantics while the structural information is not fully considered. Furthermore, the widely adopted Graph Neural Networks (GNNs) have proved their effectiveness in learning program semantics, however, they also suffer the problem of capturing the global dependencies in the constructed graph, which limits the model learning capacity. To address these challenges, in this paper, we design a novel neural network framework, named GraphSearchNet, to enable an effective and accurate source code search by jointly learning the rich semantics of both source code and natural language queries. 
Specifically, we propose to construct graphs for the source code and queries with bidirectional GGNN (BiGGNN) to capture the local structural information of the source code and queries. Furthermore, we enhance BiGGNN by utilizing the multi-head attention module to supplement the global dependencies that BiGGNN missed to improve the model learning capacity. 
The extensive experiments on Java and Python programming language from the public benchmark CodeSearchNet confirm that GraphSearchNet outperforms current state-of-the-art works by a significant margin. 
\end{abstract}

}

\maketitle
\IEEEdisplaynontitleabstractindextext
\IEEEpeerreviewmaketitle

\section{Introduction}\label{sec:intro}
With the fast development of the software industry over the past few years, the global source code over public and private repositories (\eg, on GitHub or Bitbucket) is reaching an unprecedented amount. It is already commonly recognized that the software industry is entering the ``\emph{Big Code}'' era.
Code search, which aims to search the relevant code snippets based on the natural language query from a large code corpus (\eg, Github, Stack Overflow, or private ones), has become a critical problem in the ``\emph{Big Code}'' era. In addition, some studies~\cite{liu2021opportunities, bajracharya2012analyzing} also have shown that more than 90\% of the efforts from the software developers aim at reusing the existing code. Hence, an accurate code search system can greatly improve software productivity and quality, while reducing the software development cost.

Automated code search is far from settled. Some early attempts were started by leveraging {information retrieval} (IR) techniques to capture the relationship of the code and the query by keyword matching~\cite{lu2015query, bajracharya2006sourcerer, bajracharya2012analyzing, krugler2013krugle}. However, such techniques are ad-hoc, making them limited especially when no common keywords exist in the source code and queries. Furthermore, the extracted keywords from the query tend to be short, which cannot represent the rich semantics behind the text. To address these limitations, many works expand the query format and reformulate it with different expressions~\cite{haiduc2013automatic, hill2011improving, lu2015query, mcmillan2011portfolio, lv2015codehow}. For example, Lu et al.~\cite{lu2015query} expanded the query with some synonyms generated from WordNet~\cite{miller1998wordnet} to improve the hit ratio.
CodeMatcher~\cite{liu2021codematcher} proposed an IR-based model by collecting metadata for query words to identify irrelevant/noisy ones and iteratively performing the fuzzy search with the important query words on the codebase to return the program candidates.
These IR-based techniques essentially perform the matching process based on keywords. \revised{However, since code or natural language query has different types of semantic variants, an effective code
search system requires a high-level semantic mapping between
code and natural language queries.}

To address this limitation, more recent attempts shifted to deep learning (DL)-based techniques~\cite{gu2018deep, husain2019codesearchnet, yao2019coacor, haldar-etal-2020-multi, shuai2020improving, wan2019multi, fang2021self, ishtiaq2021bert2code, du2021single}, which encode the source code and the query into vectors (\ie, learning the representations behind the program and the query). Then, the similarity between two vectors, such as cosine similarity, is computed to measure the semantic relevance between the code snippet and the query. Code snippets with a higher similarity score of the query are returned as the search results. However, most of these works only rely on sequential models \ie, Long-Short Term Memory (LSTMs)~\cite{hochreiter1997long}, Self-Attention~\cite{vaswani2017attention} to learn the vector representations for both code and query. These sequential models are struggling to learn the semantic relations because they ignore the structural information hidden in the text to learn. In addition, MMAN~\cite{wan2019multi} encoded the program sequence, abstract syntax tree (AST) and control flow graph (CFG) with LSTM~\cite{hochreiter1997long}, Tree-LSTM~\cite{mou2016convolutional} and GGNN~\cite{li2015gated} respectively on the C programming language. Then it further encoded the query sequence with another LSTM to learn the mapping relationship between the code and query, however, MMAN ignored the structural information behind the query and the limitations of GGNN \ie, the missed global dependencies in a graph~\cite{liu2020retrieval, alon2021on} in the process of GGNN learning, is not well-addressed. 
To sum up, although some recent progress has been made for automated code search, the key challenges still exist: (1) source code and the natural language queries are heterogeneous~\cite{gu2018deep}, they have completely different grammatical rules and language structure, which leads to the semantic mapping is hard; (2) the rich structural information behind in the code and the query fails to explore. Failing to utilize the rich structural information beyond the simple text may limit the effectiveness of these approaches for code search; (3) Although there are some existing works~\cite{wan2019multi} attempted to use GNNs for code search, the limitations of GNNs are not well-addressed.

To address these challenges, in this paper, we design a novel neural network framework, named \tool, towards learning the representations that fully utilize the structural information to capture the semantic relations of source code and queries for code search. In particular, since the large corpus of the query dataset is hard to collect, we follow the existing works~\cite{gu2018deep, husain2019codesearchnet, li2019neural, yan2020code} and use the summary of a code snippet for the replacement to jointly train the program encoder and the summary encoder. Specifically, we convert the program to a graph with syntactic edges (\textit{AST Edge}, \textit{NextToken} \textit{SubToken}) and data-flow edges (\textit{ComputedFrom}, \textit{LastUse} and \textit{LastWrite}) to represent the program semantics. Furthermore, we also build the summary graph based on dependency parsing~\cite{de2006generating}. For each encoder, we feed the constructed graph to Bidirectional Gated Graph Neural Network \ie, BiGGNN~\cite{chen2019reinforcement} to capture the structural information in a graph. We further enhance BiGGNN by the multi-head attention module to supplement the missed global dependencies that BiGGNN fails to learn to improve the model learning capacity. Once \tool is trained, at the query phase, given a natural language query, the summary encoder is utilized to obtain the query vector, then the top-$k$ program candidates are returned based on the cosine similarity between the query vector and the program vectors that are embedded from the program encoder on the large search code database.

To demonstrate the effectiveness of our approach, we investigate the performance of \tool against \revised{\textbf{13}} state-of-the-art baselines on Java and Python datasets from the open-sourced CodeSearchNet~\cite{husain2019codesearchnet}, which has over \textbf{2 million} functions and evaluate these approaches in terms of \textbf{5} evaluation metrics such as R@k, MRR, NDCG. We further conduct a quantitative analysis on \textbf{99} real queries to confirm the effectiveness of \tool. The extensive experimental results show that \tool significantly outperforms the baselines on the evaluation metrics. Furthermore, \tool can also produce high-quality programs based on the real natural language query from the quantitative analysis. Our code is available at \url{https://github.com/shangqing-liu/GraphSearchNet}. 
Overall, we highlight our contributions as follows:
\begin{itemize}[leftmargin=*]
    \item We propose a novel graph-based framework to capture the structural and semantic information for accurately learning the semantic mapping of the program and the query for code search. 
    \item We design \tool to improve model learning capacity by BiGGNN to capture the local structural information in a graph and multi-head attention to capture the global dependencies that BiGGNN fails to learn.
    \item We conduct an extensive evaluation to demonstrate the effectiveness of \tool on the large code corpora and the experimental results demonstrate that \tool outperforms the state-of-the-art baselines by a significant margin. we have made our code public to benefit academia and the industry.
\end{itemize}

The remainder of this paper is organized as follows: Section~\ref{sec:background} presents the background and the motivation of \tool. We elaborate our approach in Section~\ref{sec:approach}. Section~\ref{sec:setup} and Section~\ref{sec:results} are our experimental setup and the experimental results. Section~\ref{sec:discussion} gives some discussions about \tool, followed by the related works in Section~\ref{sec:related}. We conclude our paper in Section~\ref{sec:conclusion}.

\section{Background and Motivation}\label{sec:background}
In this section, we first briefly introduce the background of graph neural networks, then detail the motivation for the design of \tool and followed by introducing the existing datasets for code search and the multi-head attention that we will use in \tool.
\subsection{Graph Neural Networks}
Graph Neural Networks (GNNs)~\cite{li2015gated, hamilton2017inductive, kipf2016semi, xu2018graph2seq} have attracted wide attention over the past few years since GNNs can handle complex structural data, which contains the elements (nodes) with the relations (edges) between them. Hence, a variety of scenarios such as social networks~\cite{fan2019graph}, programs~\cite{allamanis2017learning}, chemical and biological systems~\cite{duvenaud2015convolutional} leverage GNNs to model graph-structured data. Generally, a directed graph $\mathcal G = (\mathcal  V, \mathcal E)$ contains a node list $\mathcal{V}$ and the edge list $\mathcal{E}$, where $(u, v) \in \mathcal{E}$ denotes an edge from the node $u$ to the node $v$. The learning process for GNNs is to propagate neural messages (node features) in the neighboring nodes and it is named neural message passing. It consists of multiple hops and at every hop, each node aggregates the received messages from its neighbors and updates the representation by combining the aggregated incoming messages with its own previous representation. Formally, for a node $v$ has an initial representation $\boldsymbol h^0_{v} \in \mathbb R^d$, where $d$ is the dimensional length. The initial representation is usually derived from its label or the given features. Then, a hop updates its representation by the message passing and obtains the new representation $\boldsymbol h^1_{v} \in \mathbb R^d$. This process can be expressed as follows:
\begin{equation}
    \boldsymbol h^{(k)}_{v} = f(\boldsymbol h^{(k-1)}_{v}, \{\boldsymbol h^{(k-1)}_{u} | u \in N_v\}; \theta_k)
\end{equation}
where $ N_v$ is a set of nodes that have edges to $v$. $k$ denotes $k$-th hop and the total number of hops $K$ is determined empirically as a hyper-parameter and $1 \leq k \leq K$. The function $f$ usually distinguishes different GNN variants. For example, graph convolution networks (GCN)~\cite{kipf2016semi} can be expressed as:
\begin{equation}
    \boldsymbol h^{(k)}_{v} = \sigma( \sum_{u \in {N_v \cup \{v\}}} \frac{1}{c_{v,u}}\boldsymbol W \boldsymbol h^{(k-1)}_{u})
\end{equation}
where $c_{v,u}$ is a normalization factor and it is often set to $\sqrt{|N_v|\cdot|N_u|}$ or $|N_v|$, $\boldsymbol W $ is the learnable weight and $\sigma$ is a non-linearity such as rectified linear unit (ReLU)~\cite{nair2010rectified}. Another widely used GNN variant in the program scenario is gated graph neural network (GGNN)~\cite{li2015gated}, which uses a recurrent unit $r$ \eg, GRU~\cite{cho2014learning} or LSTM~\cite{hochreiter1997long} for the update. The message passing can be calculated as follows:
\begin{equation}
    \boldsymbol h^{(k)}_{v} = r( \boldsymbol h^{(k-1)}_{v}, \sum_{u \in {N_v}} \boldsymbol W \boldsymbol h^{(k-1)}_{u}; \boldsymbol{\theta}_r)
\end{equation}
where $\boldsymbol{\theta}_r$ is the recurrent cell parameters and $\boldsymbol W $ is the learnable weight.

\subsection{Motivation}
Program semantics learning by deep learning techniques can be considered as the fundamental problem for a variety of code-related tasks. Compared with considering the program as a flat sequence of tokens with the sequential model such as LSTMs~\cite{hochreiter1997long} or Self-Attention~\cite{vaswani2017attention} to learn program semantics, Allamanis et al~\cite{allamanis2017learning} lighted up this field by converting the program into a graph with different relations of the nodes to represent the program semantics and further utilized GNNs to capture the node relations. After that, many graph-based works for different tasks~\cite{zhou2019devign, liu2020retrieval, fernandes2018structured, allamanis2020typilus, hellendoorn2019global, wan2019multi} have emerged in both AI and SE community. These works have proved the effectiveness of GNNs~\cite{kipf2016semi, li2015gated, allamanis2017learning} in modelling a program to capture program semantics. Furthermore, due to the powerful relation learning capacity of GNNs, they also have been widely used in many NLP tasks, \eg, natural question generation (QG)~\cite{chen2019reinforcement, su2020multi}, conversational machine comprehension (MC)~\cite{chen2019graphflow, song2018exploring, de2018question}. Annervaz et al.~\cite{annervaz2018learning} further confirmed that augmenting graph information with LSTM can improve the performance of many NLP tasks. Hence, inspired by these works, in this paper, we propose our approach to convert the program and the query into graphs with GNNs to learn the semantic relations for code search. 

\begin{figure}[tp]
     \centering
     \includegraphics[width=0.5\textwidth]{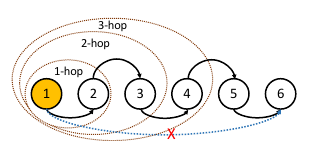}
     \caption{Message passing of GNNs within $K$-th neighborhood information where the dash area is the receptive field that node 1 knows.} 
     \label{fig:problem}
\end{figure}

\begin{figure}[tp]
     \centering
     \includegraphics[width=0.5\textwidth]{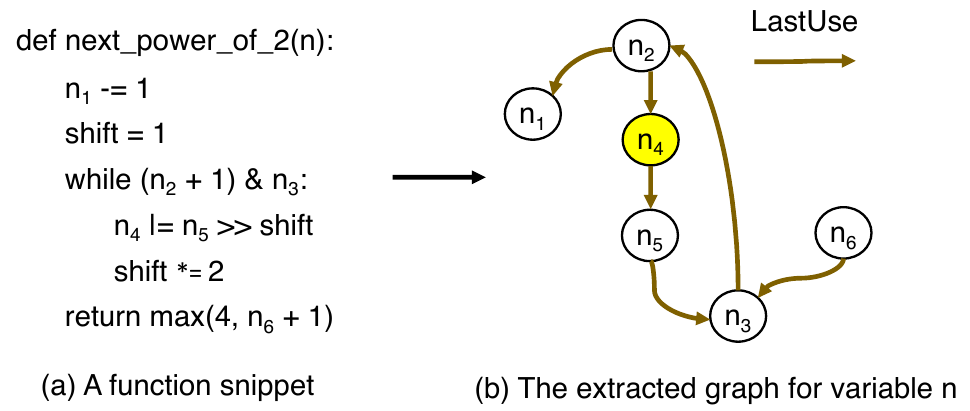}
     \caption{\revised{A real function example, please note that we distinguish the variable n into n, n\textsubscript{1}, n\textsubscript{2}, n\textsubscript{3}, n\textsubscript{4}, n\textsubscript{5} and n\textsubscript{6} for ease of presentation. Furthermore, since the constructed graph for this function is complex, we extract a subgraph that only contains the LastUse edge for the variable ``n''  for better presentation and explanation.}} 
     \label{fig:real_code_mmotivation}
\end{figure}

However, recent advanced works on GNNs~\cite{liu2020retrieval, alon2021on, hellendoorn2019global} have proved that GNNs are powerful to capture the signals from the short-range nodes, while the long-range information from the distant nodes cannot well-handled. It is caused by the message passing computation process. Specifically, the total number of hops $K$ limits the network can only sense the range of the interaction between nodes at a radius of $K$ \ie, the receptive field at $K$. We give a simple example as shown in Fig.~\ref{fig:problem} for better explanation. In particular, when $K=1$, node 1 can only know the information from its neighbor \ie, node 2. Furthermore, when $K=2$, node 1 knows the information from its neighbor \ie, node 2 and node 2's neighbor \ie, node 3. When we increase $K$ to 3, node 1 can sense the information from node 2, node 3 and node 4. Hence, to let node 1 know the information about its far-distant node, we must increase $K$ to a large value. However, the hyper-parameter $K$ can not be set to a large value due to the over-squashing~\cite{alon2021on} and over-smoothing~\cite{li2018deeper, oono2019graph} problem in GNNs. Hence, empirically, $K$ is often set to a small value~\cite{liu2020retrieval, kipf2016semi} to avoid this problem. It leads GNNs to be powerful in capturing the local structural information while failing to capture the global dependencies in a graph. 
\revised{We further provide a real code snippet to illustrate the global dependencies that GNNs fail to capture. As shown in Fig.~\ref{fig:real_code_mmotivation}a, this function is from open-source GitHub project\footnote{ \href{https://github.com/glue-viz/glue-vispy-viewers/blob/54a4351d98c1f90dfb1a557d1b447c1f57470eea/glue_vispy_viewers/extern/vispy/visuals/collections/base_collection.py\#L22-L29}{\textit{https://github.com/glue-viz/glue-vispy-viewers}}}. Since the completed constructed graph for this function is complex, we just extract a subgraph that contains the LastUse edge for the variable ``n'' for illustration, which is shown in Fig.~\ref{fig:real_code_mmotivation}b. We can observe that in this directed graph, the nodes n\textsubscript{2}, n\textsubscript{3}, n\textsubscript{4}, n\textsubscript{5} form a loop. In addition, there is a directed edge pointed from n\textsubscript{2} to n\textsubscript{1}. Hence, for a specific node n\textsubscript{4}, during the message passing in GNNs, it only requires $K=4$ (i.e., receptive field) to sense the information from n\textsubscript{1}, n\textsubscript{2},  n\textsubscript{3} and n\textsubscript{5}. We also find that n\textsubscript{4} cannot communicate with n\textsubscript{6} because they are unreachable (i.e., there is no directed path pointed from n\textsubscript{4} to n\textsubscript{6}). From Fig.~\ref{fig:real_code_mmotivation}a, we find that n\textsubscript{4} and n\textsubscript{6} are closely related in the program semantics, however, node n\textsubscript{4} cannot learn the effective features from node n\textsubscript{6} by the message passing in the graph. Hence, it is necessary to capture the global dependencies between  n\textsubscript{4} and n\textsubscript{6} to mitigate the limitation of traditional GNNs.} It inspires us to design a new neural network for code search.  

\subsection{Existing Datasets for Code Search}
There are some existing datasets designed for code search, such as CodeSearchNet~\cite{husain2019codesearchnet}, DeepCS~\cite{gu2016deep}, FB-Java~\cite{li2019neural}, CosBench~\cite{yan2020code}. The released dataset by DeepCS is processed by the authors and the raw data cannot be obtained for the graph construction. The other datasets such as FB-Java and CosBench are both collected from Java projects. In contrast, CodeSearchNet contains data from six programming languages, besides Java, it also consists of data from other programming languages such as Python. Since different programming language has specific grammatical features, \tool aims at prove the proposed approach is robust against different languages, hence we select Python and Java data from CodeSearchNet for the evaluation. Furthermore, in deep code search, a sufficient amount of (program, query) pairs is hard to collect and CodeSearchNet utilizes (program, summary) pairs for the replacement at the learning phase, where the summary describes the functionality of a function. CodeSearchNet further provides a number of 99 real natural language queries \eg, ``convert decimal to hex'' to retrieve the related programs from the search codebase, which is independent of the dataset that is used for training the model. We also follow these settings for evaluation. 

\subsection{Multi-Head Attention}\label{sec:multi-head}
Self-attention is the key idea in the transformer~\cite{vaswani2017attention}, which is widely used in NLP. An attention function can be described as mapping a query and a set of key-value pairs to an output, where the query, keys, values and output are all vectors. The particular Scaled Dot-Product Attentionr~\cite{vaswani2017attention} can be expressed as follows:
\begin{equation}
\begin{split}
    \begin{gathered}
    \mathrm{Attention(\boldsymbol{Q}, \boldsymbol{K}, \boldsymbol{V})} = \mathrm{Softmax}(\frac{\boldsymbol{Q}\boldsymbol{K}^{T}}{\sqrt{d_{k}}}) \boldsymbol{V}
    \end{gathered}
\end{split}
\end{equation}
where $d_{k}$ is the dimensional length, $\boldsymbol{Q}$, $\boldsymbol{K}$, and $\boldsymbol{V}$ represent the query, key, and value matrices. To further improve the expression capacity of the self-attention, it is beneficial to linearly project the queries, keys and values $h$ times with different linear projections and then concatenate and project to obtain the final outputs and this calculation process is named multi-head attention~\cite{vaswani2017attention}:
\begin{equation}~\label{eq:muti-head}
\begin{split}
    \begin{gathered}
    \mathrm{MultiHead}(\boldsymbol{Q}, \boldsymbol{K}, \boldsymbol{V})=\mathrm{Concat}(\mathrm{head}_1, ..., \mathrm{head}_{h}) \boldsymbol{W}^{O} \\
    \mathrm{where} \; \mathrm{head}_{i} = \mathrm{Attention(\boldsymbol Q \boldsymbol{W}^{Q}_{i}, \boldsymbol K \boldsymbol{W}^{K}_{i}, \boldsymbol V \boldsymbol{W}^{V}_{i})}
    \end{gathered}
\end{split}
\end{equation}
\noindent where $\boldsymbol{W}^{Q}_{i} \in \mathbb{R}^{d_{model} \times d_{k}}$, $\boldsymbol{W}^{K}_{i} \in \mathbb{R}^{d_{model} \times d_{k}}$, $\boldsymbol{W}^{V}_{i} \in \mathbb{R}^{d_{model} \times d_{v}}$ and $\boldsymbol{W}^{O} \in \mathbb{R}^{hd_v \times d_{model}}$ are model parameters and $h$ is the number of heads.

\begin{figure*}[tp]
     \centering
     \includegraphics[width=1.0\textwidth]{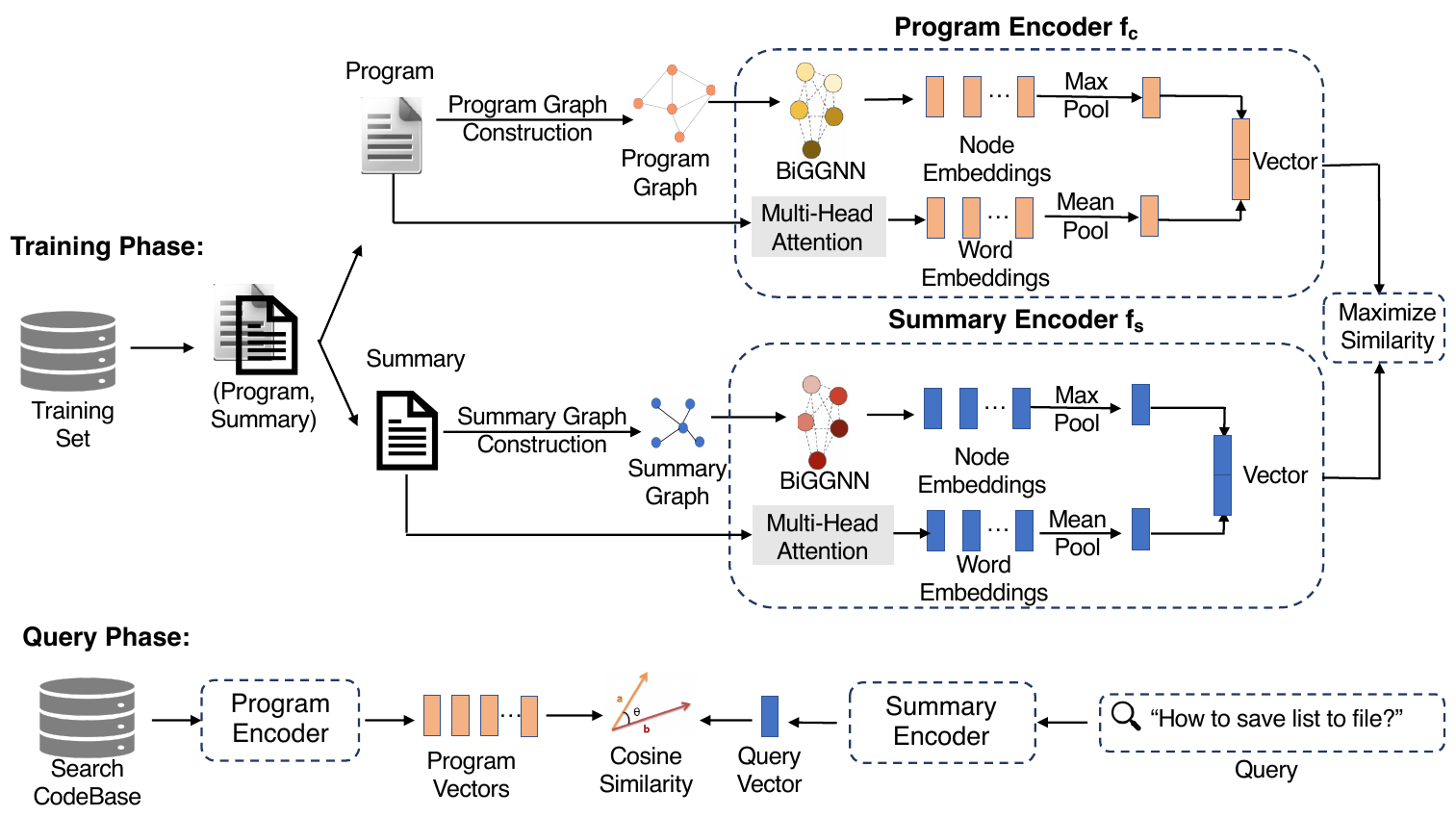}
     \caption{The framework of \tool.} 
     \label{fig:framework}
\end{figure*}

\section{GraphSearchNet}\label{sec:approach}
In this section, we introduce our framework \tool, as shown in Fig.~\ref{fig:framework}, which includes three sequential parts: 1) \emph{Graph Construction}, which constructs directed graphs for programs and summaries with comprehensive semantics. 2) \emph{Training Phase}, which jointly learns two separate encoders i.e., the program encoder $f_c$ and the summary encoder $f_s$ to obtain the vector representations respectively. Each encoder includes two modules: BiGGNN, which aims at capturing local graph structural information, and multi-head attention, which supplements the global dependencies that GNNs missed to enhance the model learning capacity. 3) \emph{Query Phase}, which returns a set of top-$k$ candidates from the search codebase that are most similar to the given query based on the cosine similarity, where the query is independent with the summary used for model training and the search codebase is different from the training set that used to train the model.




\subsection{Problem Formulation}
The goal of code search is to find the most relevant program fragment $c$ based on the query $q$ in natural language. Formally, given a set of training data $D=\{(c_i, q_i) | c_i \in \mathcal{C}, q_i \in \mathcal{Q}\}, i\in\{1,2,...,n\}$, where $\mathcal{C}$ and $\mathcal{Q}$ denote the set of functions and the corresponding queries. We define the learning problem as:
\begin{equation}
    \min \sum_{i=1}^n \mathcal{L}(f_c(c_i), f_q(q_i))
\end{equation}
where $f_c$ and $f_q$ are the separate encoders \ie, the neural network for programs and queries, which are learnt from the training data $D$. However, in the real scenario, since $D$ is hard to collect, it is a common practise~\cite{gu2018deep, husain2019codesearchnet, li2019neural, yan2020code} to replace $q_i$ with the summary $s_i$ regarding the function $c_i$ i.e., $D=\{(c_i, s_i) | c_i \in \mathcal{C}, s_i \in \mathcal{S}\}$ where $\mathcal{S}$ is the corresponding summary set. The summary $s_i$ usually describes the functionality of the function $c_i$. Once the encoders are trained \ie, $f_c$ and $f_s$, given a query $q$ and the search code database $\mathcal{C}_{base}$ where  $\mathcal{C}_{base} \neq D$ , we can obtain the most relevant program as:
\begin{equation}
c = \max_{ c_i \in \mathcal{C}_{base}}\mathrm{sim}(f_c(c_i), f_s(q))
\end{equation}
where $\mathrm{sim}$ is a function such as the cosine similarity function to measure the semantic similarity between the program vector and query vector, which are produced by the encoder $f_c$ and $f_s$ respectively. 



\begin{figure*}[tp]
     \centering
     \includegraphics[width=1.0\textwidth]{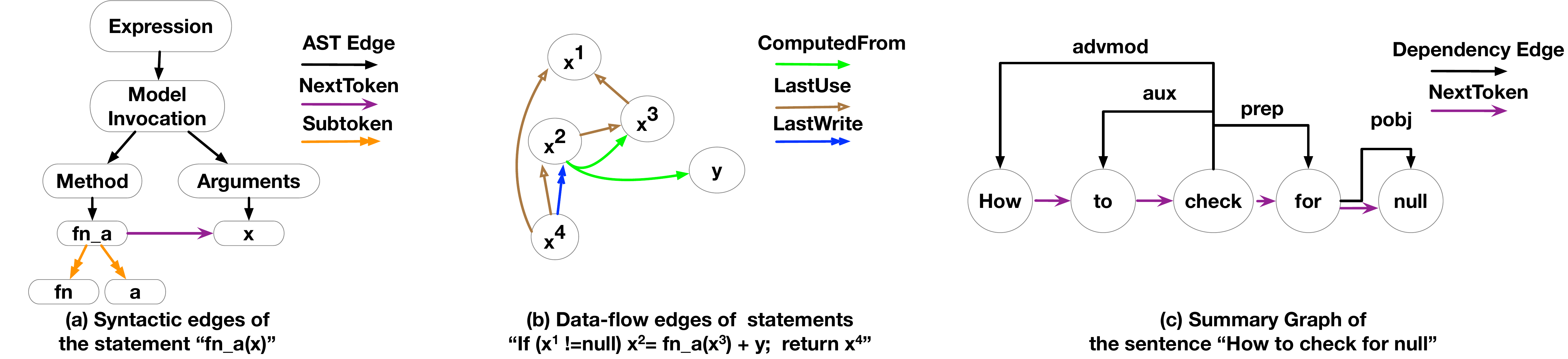}
     \caption{An example of the constructed graphs where (a) and (b) are the syntactic edges and data-flow edges of the program graph with a simple program snippet, please note that we distinguish the variable x into
     x\textsuperscript{1}, x\textsuperscript{2}, x\textsuperscript{3}, x\textsuperscript{4} for ease of clarity in data-flow edges, (c) is the constructed summary graph.} 
     \label{fig:graph}
\end{figure*}

\subsection{Graph Construction}
We introduce the graph construction for both programs and summaries that will be used for the encoder to learn.
\subsubsection{Program Graph Construction}
Given a program $c$, we follow Allamanis et al.~\cite{allamanis2017learning} and extract its multi-edged directed graph $\mathcal G = (\mathcal  V, \mathcal E)$, where $\mathcal V$ is a set of nodes that are built on the Abstract Syntax Tree (AST) and $\mathcal E$ is the edges that represent the relationships between nodes. In particular, AST consists of terminal nodes and non-terminal nodes where terminal nodes correspond to the identifiers in the program and the non-terminal nodes represent different compilation units such as ``Assign'', ``BinOp'', ``Expr''. The edges can be categorized into syntactic edges \ie, \textit{AST Edge}, \textit{NextToken} and data-flow edges \ie, \textit{ComputedFrom}, \textit{LastUse}, \textit{LastWrite}. In addition, according to Cvitkovic et al.~\cite{cvitkovic2019open}, \textit{SubToken} edge can further enrich the semantics of a program on the graph, we also include it and introduce extra subtoken nodes appearing in the identifier of the terminal node and connecting them to their original nodes. Hence, the node set $\mathcal V$ is the union of the entire AST nodes and subtoken nodes. 
The details of these types of edges with a simplified example of our constructed program graph in Fig.~\ref{fig:graph}a and Fig.~\ref{fig:graph}b are presented as follows:
\begin{itemize}[leftmargin=*]
    \item \textit{AST Edge}, connects the entire AST nodes based on the abstract syntax tree.
    \item \textit{NextToken}, connects each leaf node in AST to its successor. As shown in Fig.~\ref{fig:graph}a, for the statement ``fn\_a(x)", there is a \textit{NextToken} edge points from ``fn\_a'' to ``x''.
    \item \textit{SubToken}, defines the connection of subtokens split from a identifier \ie, variable names, function names based on \textit{camelCase} and \textit{pascal\_case} convention. For example, ``fn\_a'' will be divided into ``fn'' and ``a''. We further introduce the extra subtoken nodes apart from AST nodes to describe this relation.
    \item \textit{LastUse}, represents the immediate last read of each occurrence of variables. As shown in Fig.~\ref{fig:graph}b, x\textsuperscript{3} points to x\textsuperscript{1} since x\textsuperscript{1} is used as the conditional judgement, x\textsuperscript{4} points to x\textsuperscript{1} and x\textsuperscript{2} because if the conditional judgement is not satisfied, x\textsuperscript{4} points to x\textsuperscript{1} directly, otherwise it will point to x\textsuperscript{2}.
    \item \textit{LastWrite}, represents the immediate last write of each occurrence of variables. Since there is an assignment statement to update x\textsuperscript{2}, x\textsuperscript{4} points to x\textsuperscript{2} with a \textit{LastWrite} edge.
    \item \textit{ComputedFrom}, connects the left variable to all right variables that appeared in an assignment statement. In Fig.~\ref{fig:graph}b, x\textsuperscript{2} connects x\textsuperscript{3} and y by \textit{ComputedFrom} edge.
\end{itemize}

\subsubsection{Summary Graph Construction}
Constituency parse tree~\cite{jurafskyspeech} and dependency parse tree~\cite{de2006generating} are widely used to extract the structural information for the natural language text~\cite{chen2019reinforcement}. However, compared with constituency parse tree, dependency parse tree follows the dependency grammars~\cite{keselj2009speech} and describes the dependencies by the directed linked edges between tokens in a sentence, hence the constructed graph describes the relations among the tokens without redundant information \ie, no non-terminal nodes in the constructed graph. Motivated by this, we also construct a directed graph based on dependency parsing~\cite{de2006generating} for the summary. Specifically, given a summary $s$, we extract its dependency parsing graph as $\mathcal G' = (\mathcal  V', \mathcal E')$, where $ \mathcal V'$ is a set of nodes where each node is the token in the original sequence $s$, $\mathcal E'$ denotes the relations between these tokens. Different edges represent tokens with different grammatical relations. Furthermore, there are a total of 49 different dependency edge types~\cite{de2008stanford},  
for example, the edge ``prep'' is a prepositional modifier of a verb, adjective, or noun that serves to modify the meaning of the verb, adjective, noun~\cite{de2008stanford}. 
In addition, we also construct \textit{NextToken} and \textit{SubToken} edge in the summary graph, which is similar to the program graph. 
To summarize, given a query ``How to check for null'', which is shown in Fig.~\ref{fig:graph}c, we can obtain its summary graph where each token has different dependency relations with other tokens, for example, ``How'', ``to'' and ``for'' are used to describe the relations with the verb ``check'' in the relations of `advmod'', ``aux'', ``prep'' respectively.

\subsection{Encoder}
By the program graph construction and summary graph construction, we can get the program $c$ with its program graph $\mathcal G$ and the summary $s$ with its summary graph $\mathcal G'$, we further feed them into two encoders $f_c$ and $f_s$ to learn the vector representations for the program and the summary separately. A variety of GNN variants are proposed such as GCN~\cite{kipf2016semi}, GGNN~\cite{li2015gated}, GIN~\cite{xu2018powerful} to model the graph-structured data. 
Considering that most existing GNN variants ignore the direction for the message passing and treat the graph as the undirected graph, in this work, inspired by the recent advanced Bidirectional Gated Graph Neural Network (BiGGNN)~\cite{chen2019reinforcement}, which leverages both incoming and outgoing message passing for the node interaction, we also use it for the graph learning. Furthermore, to supplement the global dependencies that are missed in GNNs, we propose a multi-head attention module to further improve the expression of BiGGNN. In the following, we only use the program $c$ with its constructed program graph $\mathcal G$ for the explanation, the operations for the summary encoder $f_s$ are the same with the program encoder $f_c$, the difference is that we feed the summary $s$ with its constructed graph $\mathcal G'$ for the summary encoder to learn the representation.
\subsubsection{Bidirectional GGNN}
Given a program graph $\mathcal G = (\mathcal  V, \mathcal E)$, BiGGNN learns the node embeddings from both incoming and outgoing directions for the graph. In particular, each node $v \in \mathcal V$ is initialized by a learnable embedding matrix $\boldsymbol E$ and gets its initial representation $\boldsymbol h_v^0 \in \mathbb{R}^{d}$ where $d$ is the dimensional length. We apply the message passing function for a fixed number of hops i.e., $K$. At each hop $k \leq K$, for the node $v$, we apply a summation aggregation function to take as input a set of incoming (or outgoing) neighboring node vectors and outputs a backward (or forward) aggregation vector. The message passing is calculated as follows, where $N_{(v)}$ denotes the neighbors of node $v$ and $\dashv/\vdash$ is the backward and forward direction.
\begin{equation}
\begin{aligned}
  \boldsymbol h_{\mathcal{N}_{\dashv(v)}}^k = \mathrm{SUM}(\{\boldsymbol h_u^{k-1}, \forall u \in \mathcal{N}_{\dashv(v)} \}) \\
  \boldsymbol h_{\mathcal{N}_{\vdash(v)}}^k = \mathrm{SUM}(\{\boldsymbol h_u^{k-1}, \forall u \in \mathcal{N}_{\vdash(v)} \}) 
 \end{aligned}
\end{equation}
Then, we fuse the node embedding for both directions:
\begin{equation}
    \boldsymbol h_{\mathcal{N}_{(v)}}^k = \mathrm{Fuse}(\boldsymbol h_{\mathcal{N}_{\dashv(v)}}^k, \boldsymbol h_{\mathcal{N}_{\vdash(v)}}^k)
\end{equation}
 Here the fusion function is designed as a gated sum of two inputs.
 \begin{equation}
 \label{func:fusion}
 \begin{gathered}
 \mathrm{Fuse(\boldsymbol a,\boldsymbol b)} = \boldsymbol z \odot \boldsymbol a + (1-\boldsymbol z) \odot \boldsymbol b \\ \boldsymbol z = \sigma(\boldsymbol W_z[\boldsymbol a;\boldsymbol b;\boldsymbol a \odot \boldsymbol b; \boldsymbol a- \boldsymbol b ] +\boldsymbol b_z)
 \end{gathered}
  \end{equation}
 where $\odot$ is the component-wise multiplication, $\sigma$ is a sigmoid function and $\boldsymbol z$ is a gating vector. Finally, we feed the resulting vector to a Gated Recurrent Unit (GRU)~\cite{cho2014learning} to update node representations.
 \begin{equation}
 \boldsymbol h_v^k =\mathrm{GRU}(\boldsymbol h_v^{k-1}, \boldsymbol h_{\mathcal{N}_{(v)}}^k)
 \end{equation}
 After $K$ hops of computation, we obtain the final node representation $\boldsymbol h_v^K$ and then apply max-pooling over all nodes $\{\boldsymbol h_v^K, \forall v \in  V \}$ to get a $d$-dim graph representation $\boldsymbol h^g$. 
 \begin{equation}
 \label{eq:maxpool}
 \boldsymbol h^g =\mathrm{maxpool}(\mathrm{FC}(\{\boldsymbol h_v^K, \forall v \in  V \}))
 \end{equation}
 where FC is a fully-connected layer.
 
\subsubsection{Multi-Head Attention}
To further capture the global interactions in the graph that BiGGNN missed~\cite{alon2021on, liu2020retrieval}, we design a multi-head attention module to improve the model learning capacity. Note that in \tool, the constructed program graph consists of terminal/non-terminal nodes and the terminal nodes are the tokens/subtokens of the original program $c$, hence we directly employ  multi-head attention~\cite{vaswani2017attention} over the sequence \ie, $c$ to capture the global dependency relations among these terminal nodes while ignoring the non-terminal nodes for effective learning. Specifically, given the query\footnote{The query matrix is different with the natural language query, where the former is a matrix used in self-attention, and the latter is used to return the programs in code search. More explanations about the query matrix can be found in Section~\ref{sec:multi-head}.}, key and value matrix defined as $\boldsymbol Q$, $\boldsymbol K$ and $\boldsymbol V$, where $\boldsymbol Q, \boldsymbol K, \boldsymbol V = (\boldsymbol {E}_{t_1},...,\boldsymbol{E}_{t_l})$ are the embedding matrices for the program $c$ and $t_i$ is the $i$-th token in $c=\{t_1, ..., t_l\}$, the output by multi-head attention can be expressed as follows:
\begin{equation}\label{eq:initialbilstm}
\boldsymbol h_{t_1},...,\boldsymbol h_{t_l} = \mathrm{MultiHead}(\boldsymbol Q, \boldsymbol K, \boldsymbol V)
\end{equation}
To get the final representation vector $\boldsymbol h^c$ over $c$, we use the mean pool operation on $\boldsymbol h_{t_1},...,\boldsymbol h_{t_l}$ as follows.
 \begin{equation}
 \boldsymbol h^c =\mathrm{meanpool}(\{\boldsymbol h_{t_i}, \forall t_i \in c \})
 \end{equation}
Here we choose the meanpool operation rather than the maxpool operation that used in BiGGNN is based on our experiments and we found that meanpool operation can get higher results in multi-head attention. 
Finally, we concatenate the vectors produced from Bidirectional GGNN i.e., $\boldsymbol h^g$ and multi-head attention i.e., $\boldsymbol h^c$ to get the final representation $\boldsymbol r = [\boldsymbol h^g; \boldsymbol h^c]$ for the program.

By the program encoder $f_c$ and the summary encoder $f_s$, we can obtain the paired vector defined as $(\boldsymbol r, \boldsymbol r')$ where $\boldsymbol r$ and $\boldsymbol r'$ are the output vectors for the program and the summary respectively.

\subsection{Training}
Two different loss functions are used to train a model in deep code search systems~\cite{gu2016deep, husain2019codesearchnet}. The ranking loss~\cite{collobert2011natural, frome2013devise} that used in Gu et al.~\cite{gu2016deep} needs to determine a hyper-parameter $\epsilon$, however by our preliminary experiments, we found that the value is hard to determine manually, although it was set to 0.05 in their experiments. Instead, following CodeSearchNet~\cite{husain2019codesearchnet}, we directly use Cross-Entropy for the optimizing. Specifically, given $n$ pairs of $(c_i, s_i)$, we train $f_c$ and $f_s$ simultaneously by minimizing the loss as follows:
\begin{equation}
    -\frac{1}{n} \sum_{i}^{n} \mathrm{log} \left( \frac{\boldsymbol r_i^T \boldsymbol r'_i}{\sum_{j} \boldsymbol r_j^T \boldsymbol r_i'} \right)
\end{equation}
The loss function minimizes the inner product of $(\boldsymbol r'_{i}, \boldsymbol r_{i})$ between the summary $s_i$ and its corresponding program $c_i$, while maximizing the inner product between the summary $s_i$ and other distractor programs, \ie, $c_j$ where $ j \neq i$.

\subsection{Code Searching}
Once the model is trained, given a query $q$, \tool aims to return a set of the most relevant programs. To achieve this, we first embed and stored all programs into vectors by the learnt program encoder $f_c$ in an offline manner and these programs are from the search code database $\mathcal{C}_{base}$, which is independent with the data set used for model training. At the online query phase, for a new query $q$, \tool embeds it via the summary encoder $f_s$ and computes the cosine similarity between the query vector with all stored program vectors in the search codebase. The top-$k$ programs whose vectors are the most similar \ie, the highest top-$k$ similarity values, to the query $q$ are returned as the results. In \tool, we select 10 candidate programs \ie, $k=10$ as the returned results.


\section{Experimental Setup}\label{sec:setup}
In this section, we introduce the experimental setup including the dataset, evaluation metrics, compared baselines and the hyper-parameter configuration of \tool.
\subsection{Dataset}
We select Java and Python data from CodeSearchNet~\cite{husain2019codesearchnet} to evaluate our approach that can be generalized to different programming languages. The reason to choose Java and Python datasets is that there are some existing open-sourced tools~\cite{features-javac,fernandes2018structured} for Java and Python programming language to facilitate constructing the program graph. We follow the same train-validation-test split with CodeSearchNet.
In addition, we utilize spaCy~\cite{spacy2} to construct the summary graph.
At the query phase, for all programs in the search codebase, we employ the open-sourced ElasticSearch~\cite{gormley2015elasticsearch} to store the vectors obtained by the learnt program encoder $f_c$ for acceleration. Converting all programs from the search codebase into vectors costs nearly 70 minutes for Java or Python over \textbf{756k} samples in an offline manner. However, the query process provided by ElasticSearch is fast and it only takes about 0.15 seconds to produce 10 candidates for each query based on the cosine similarity. To sum up, in the evaluation section, we first investigate the performance of \tool on the test set with some automatic metrics. Then we conduct a quantitative analysis over the real 99 queries with the programs from the search codebase to confirm the effectiveness of our approach.

\subsection{Automatic Evaluation Metrics}\label{sec:automaitc-metrics}
Similar to the previous works~\cite{gu2018deep, haldar-etal-2020-multi, shuai2020improving}, we use SuccessRate@$k$, Mean Reciprocal Rank (MRR) as our automatic evaluation metrics. We further add Normalized Discounted Cumulative Gain (NDCG)~\cite{husain2019codesearchnet} as another evaluation metric. 
\begin{itemize}[leftmargin=*]
 \item \textbf{SuccessRate@$k$.} SuccessRate@$k$ measures the percentage of the correct results that existed in the top-$k$ ranked results and it can be calculated as follows:
\begin{equation}
   \mathrm{SuccessRate@k} = \frac{1}{|Q|} \sum_{q=1}^{|Q|} \delta(\mathrm{FRank}_q \leq k)
\end{equation}
where $Q$ is a set of queries, $\delta(\cdot)$ is a function, which returns 1 if the input is true and 0 otherwise. FRank is the rank position of the first hit result in the result list and we set $k$ to 1, 5, 10. 
\item \textbf{MRR.} MRR is the average of the reciprocal ranks of results for a set of queries $Q$. The reciprocal rank of a query is the inverse of the rank of the first hit result, defined as follows:
\begin{equation}
   \mathrm{MRR} = \frac{1}{|Q|} \sum_{q=1}^{|Q|} \frac{1}{\mathrm{FRank}_q}
\end{equation}
\item \textbf{NDCG.} NDCG measures the gain of the result based on its position in the result list and it is the division of discounted cumulative gain (DCG) and ideal discounted cumulative gain (IDCG) where DCG and IDCG are calculated as follows:
\begin{equation}
\begin{aligned} 
   \mathrm{DCG_p} = \sum_{i=1}^{p} \frac{2^{\mathrm{rel}_i} - 1}{\mathrm{log_2}(i+1)} && \mathrm{IDCG_p} = \sum_{i=1}^{\mathrm{|REL|}_p} \frac{2^{\mathrm{rel}_i} - 1}{\mathrm{log_2}(i+1)}
\end{aligned}
\end{equation}
where $p$ is the rank position, $\mathrm{rel}_i$ is the graded relevance of the result at position $i$ and $\mathrm{|REL|}_p$ is the list of relevant results up to position $p$. We set $p$ equal to 10 for all experiments. 
\end{itemize}
To sum up, for the above automatic metrics, the higher values, the better performance the approach achieves.

\subsection{Compared Baselines}
To demonstrate the effectiveness of \tool, we select the following state-of-the-art approaches as our baselines, which are summarized as follows.
\begin{itemize}[leftmargin=*]
    \item \textbf{CodeSearchNet}~\cite{husain2019codesearchnet}. CodeSearchNet provided a framework to encode programs and their summaries for code search. It consists of different encoders, \ie, Natural Bag of Words (NBoW), 1D Convolutional Neural Network (1D-CNN), Bidirectional RNN (biRNN) and Self-Attention (SelfAtt) for encoding. Specifically, SelfAtt encoder has 3 identical multi-head attention layers with 128-dimensional length and each layer has 8 heads to learn different subspace features.
     \item \textbf{UNIF}~\cite{cambronero2019deep}. UNIF followed the framework of CodeSearchNet~\cite{husain2019codesearchnet}, but it replaced the self-attention mechanism by combining each bag of code token vectors into a single code vector for the program and averaging a bag of query token vectors as a single vector for a query. Then it retrieved the most similar programs on the query.
      \item \textbf{DeepCS}~\cite{gu2018deep}. DeepCS jointly learnt code snippets and natural language descriptions into a high-dimensional vector space. For code encoding, it fused method name, API sequence, and sequence tokens to learn the program vector. For the description, another RNN was utilized for learning the embedding. Based on the learnt vectors, it further used the ranking loss function~\cite{collobert2011natural, frome2013devise} to train the model and retrieved the most similar programs.
      \item \textbf{CARLCS-CNN}~\cite{shuai2020improving}. CARLCS-CNN embedded the representations for the code and query via a co-attention mechanism and co-attended the semantic relations of the embedded code and query via row/column-wise max-pooling. Then the learnt vectors were used for code search.
      \item \textbf{TabCS}~\cite{xu2021two}. TabCS incorporated the code textual features \ie, method name, API sequence and tokens, the code structural feature \ie, AST and the query feature \ie, tokens with a two-stage attention network to learn the better code and query representation. Then two datasets were used for the evaluation. One of them is the Java dataset from CodeSearchNet~\cite{husain2019codesearchnet}, which is the same as us.
      \item \textbf{QC-based CR}~\cite{yao2019coacor}. The original paper first designed a code annotations model to take the code as an input and output an annotation. Then it utilized the generated annotation with the query for a QN-based code retrieval model. In addition, they also added a QC-based code retrieval model to take the code with its query as the input to distinguish the code snippets from others. The entire model cannot run correctly by the official code, for simplicity, we only select QC-based code retrieval model as our baseline.
      \item \revised{\textbf{Code Summarization}. We add a code summarization baseline to validate whether the generated summaries by some code summarization systems with simple query strategies can produce comparable results. Specifically, we use CodeBERT~\cite{feng2020codebert} to fine tune two code summarization models for Java and Python respectively on our dataset with the official code from CodeXGLUE~\cite{lu2021codexglue}. Then, we utilize the trained models to generate summaries for the samples from the testsets we used. Based on the generated summaries with the ground-truth summaries, we further utilize a word2vec model trained from the search codebase for vectorization. Then, we calculate the evaluation metrics based on the cosine similarity between two vectors and report the values for comparison.}
      \item \revised{\textbf{CodeMatcher~\cite{liu2021codematcher}}. CodeMatcher is a retrieval-based approach for code search. It first collected the metadata for query words to filter out the irrelevant noises and then iteratively performed the fuzzy search with the important words on the codebase. Finally, it produced a set of returned candidate programs based on the importance of tokens in the candidate code snippet that matched the important words in a query.}
\end{itemize}
We also select two widely used GNN variants \ie, GGNN and GCN to investigate the performance of other GNNs as compared to BiGGNN. We directly replace BiGGNN module with GCN or GGNN for both encoders \ie, $f_c$ and $f_s$ and keep the other settings unchanged for a fair comparison. For both GNN variants, since they are not used in the code search currently, we implement them from the scratch; for TabCS, since the evaluation dataset \ie, Java dataset from CodeSearchNet, and the evaluation metrics are the same as us, we directly take the values reported in their paper~\cite{xu2021two} for comparison, for other baselines with the released source code, we directly reproduce their methods with the default settings on our dataset. We do not compare with MMAN~\cite{wan2019multi} because their approach was conducted on C dataset and the relevant tool to construct the program graph for C programming language is not available so far. 


\subsection{Model Settings}\label{sec:settings}
We embed the most frequent 150,000 tokens for the programs and summaries in the training set with a 128-dimensional size. We set the dropout as 0.3 after the word embedding layer for the learning process. We use Adam~\cite{kingma2014adam} optimizer with an initial learning rate of 0.01 and reduce the learning rate by a factor of 0.5 with the patience equals to 2. We set the number of the maximum epoch as 100 and stop the training process when no improvement on MRR for 10 epochs. We clip the gradient at length 10. The batch size is set as 1,000, following the existing work~\cite{husain2019codesearchnet}. 1000 samples in a batch can be explained as there is a corrected program $c_i$ for the current summary $s_i$ and the remaining 999 programs in this batch are distractors for the current $s_i$ to distinguish. The number of hops on Java and Python is set to 4 and 3, respectively. The number of heads in multi-head attention is set to 2 on both Java and Python datasets for efficiency. All experiments were run on the DGX server with 80 cores and 180G RAM. Four 32 GB Nvidia Graphics Tesla V100 were used to train the models and the training process took nearly 4 hours to complete. All hyper-parameters are tuned on the validation set.

\section{Experimental Results}\label{sec:results}
In this section, we aim to answer the following research questions by our experiments:
\begin{itemize}[leftmargin=*]
    \item \textbf{RQ1:} Can \tool outperform current state-of-the-art baselines in terms of automatic metrics? 
    \item \textbf{RQ2:} What is the performance of each component used in \tool, are BiGGNN and multi-head attention both beneficial in improving the performance?
    \item \textbf{RQ3:} What is the impact of setting different values of hops and heads on the performance? Whether these values perform consistently on Java and Python datasets?
    \item \textbf{RQ4:} Can \tool provide more accurate programs for the real 99 queries provided by CodeSearchNet~\cite{husain2019codesearchnet} compared with other baselines?
\end{itemize}

\begin{table*}[t]
\centering
\caption{Experimental results on Java and Python datasets as compared to baselines, where the marker * denotes the values are taken from the original paper and the marker - denotes the unreported metrics.}
\label{table-baselines}
\scriptsize {\addtolength{\tabcolsep}{0pt}
\begin{tabular}{c|ccccc|ccccc|c}
\hline
\multicolumn{1}{c|}{\multirow{2}{*}{Model}} & \multicolumn{5}{c|}{Java}        & \multicolumn{5}{c|}{Python}    & \multicolumn{1}{c}{\multirow{2}{*}{\revised{p-value}}}   \\ \cline{2-11} 
\multicolumn{1}{c|}{}                       & R@1 & R@5 & R@10 & NDCG & MRR & R@1 & R@5 & R@10 & NDCG & MRR \\ \hline
\revised{CodeMatcher}                                      &  \revised{53.77}    &   \revised{74.63}  & \revised{80.30}   & \revised{61.67}   & \revised{62.71}   &  \revised{58.74}   &   \revised{79.46}  &   \revised{85.03}   & \revised{67.32}    &  \revised{68.47}  & \revised{0.15}\\ \hline
\revised{Summarization}                                      &  \revised{1.01}    &   \revised{3.14}  & \revised{12.41}   & \revised{4.82}   & \revised{7.77}   &  \revised{2.13}   &   \revised{17.30}  &   \revised{ 40.31}   & \revised{23.65}    &  \revised{17.05} & \revised{0.00}  \\
NBoW                                      &  52.11    &  72.65    &  78.98     &    54.30  &  61.60   &  56.85    &   78.47  &   84.21    &  55.18    &  66.60   & \revised{0.06}\\
biRNN                                        &  45.65    & 67.70    &   75.33    &  62.11    &  55.90   &  54.18    &   77.01   &   83.29    &   70.76   &    65.50 & \revised{0.05}\\
SelfAtt Encoder                                        &  42.09    &   62.96   &   71.01    &   67.18   &  52.00   &  57.81    &  79.02    &    84.46   &   {76.62}    &   67.40 & \revised{0.07} \\
1D-CNN                                        &   37.91   &   58.79   &   67.10    &   54.89  &   47.80   &  41.21  &   64.65   &  72.55    &   63.56   &   52.10 & \revised{0.00} \\
UNIF                                      &   52.79   &   73.39   &   79.67    &   46.70   &  62.20   &   57.47   &  78.95   &  84.43     &  44.25    &   67.10  & \revised{0.05}\\
DeepCS                                       &   33.40   &   56.50   &    67.30   &   48.78   &   44.64  &  53.60   &   73.40  &    78.90   &   66.10   &  62.91   & \revised{0.01}\\
CARLCS-CNN                                  &   42.60   &   57.90   &   67.50    &   47.90   &   43.31  &   68.70   &   78.10   &   82.80   &   70.08   &   67.00  & \revised{0.03}\\
TabCS & 54.70* & 68.30* & 74.80* & - &53.90* & - &-&-&-&- & \revised{0.19}\\
QC-based CR                                 &   19.03   &  40.77    &   51.68   &  33.94    &  29.72   &   21.02   &  43.83    &   54.17    &  36.26    &   32.03  & \revised{0.00}\\  \hline
GGNN                                 &   48.59   &  70.39    &   77.66   &  63.09   &  58.72   &   59.69  &  80.41    &  85.80   &  72.78    &   69.07 & \revised{0.14}\\
GCN                                 &   41.55   &  66.07    &   72.92   &  57.54    &  52.96   &   48.71   &  72.25    &  79.19    &  63.77    &  59.50  & \revised{0.01}\\ \hline
GraphSearchNet                              &   \textbf{56.99}   &   \textbf{76.03}   &    \textbf{81.15}   &   \textbf{69.47}   &   \textbf{65.80}  &    \textbf{65.31}  &   \textbf{84.21}   &  \textbf{89.05}     &  \textbf{77.30}    &  \textbf{73.90} & -    \\ \hline
\end{tabular}
}
\end{table*}

\subsection{RQ1: Compared with Other Baselines.}
Table~\ref{table-baselines} summarizes the results of \tool in line with the baseline methods. Specifically, the columns R@1, R@5 and R@10 are the results of SuccessRate@$k$, where k is 1, 5 and 10. The second row is the retrieval-based approach. The third row presents the results by considering the program and the summary as the sequential text with different networks for retrieving, the fourth row is the results for other GNN variants. \revised{From Table~\ref{table-baselines}, we find that the performance of CodeMatcher is better than DeepCS and UNIF, which is consistent with the conclusion from its original paper~\cite{liu2021codematcher}. But its performance is still worse than \tool which indicates that extracting hidden structures from programs and queries can help the model learn better semantic mapping relations.} We can observe that in the third row, the performance of different approaches is inconsistent on different programming languages, for example, in terms of MRR, on the Java dataset, UNIF gets the best performance, while SelfAtt Encoder performs the best on Python dataset. One reason we conjecture is that these sequential approaches cannot capture the semantic mappings for both program and summary, hence they may bias to some specific programming languages and cannot generate well on others. Furthermore, we also find that QC-based CR has the worst performance on both datasets, we believe it is due to the fact that we only employ part of the officially released code for comparison, however, the entire project cannot run correctly by our huge efforts. \revised{From Table~\ref{table-baselines}, we observe that using the simple query strategy to match the generated summary produced by the code summarization system with the ground truth has poor performance. We investigate the main reason for it. Since the smoothed BLEU-4 (a metric to evaluate the text similarity between the generated text with the ground-truth) on our Java and Python testset is 17.11 and 19.68, which are near to the reported values (17.65 and 19.06) from CodeXGLUE~\cite{lu2021codexglue}. It demonstrates that our trained models are correct. Hence, we believe that it is because current code summarization models are still not able to generate summaries that are highly similar with ground-truths. These reported values from CodeXGLUE for code summarization models are still lower for the real application.}
In addition, we find that GGNN outperforms GCN by a significant margin for code search, which is reasonable since GRU cell in GGNN has been proven the effectiveness in filtering unnecessary features than GCN~\cite{alon2021on}. This also explains why current state-of-the-art models~\cite{allamanis2017learning, allamanis2020typilus, zhou2019devign} select GGNN as the basic block.

The last row shows our results, we can see that \tool outperforms these sequential approaches (the second row) by a significant margin on both Java and Python datasets, indicating that \tool can return more relevant programs. The results indicate that, compared with these sequential models, which treat programs and summaries as sequences, by capturing the structural information, \tool could learn the semantic relations better, making it more effective and accurate in code search. Furthermore, we find that \tool has a better performance than GGNN and GCN, we attribute it to the bidirectional message passing and multi-head attention module that can have powerful learning capacity. \revised{We also calculate the p-value of each baseline and \tool on both Java and Python dataset. The values are shown in the rightmost column of Table~\ref{table-baselines} where the value of 0.00 denotes that the p-value is too small to be close to zero. We find that there are 10 baselines (13 baselines in total) that p-values are lower than 0.1, which demonstrates that \tool provides statistically significant improvements compared with these baselines.}
More details about the performance of each module can be found that Section~\ref{sec:ablation}. \revised{Finally, we can see that the overall performance on Python is superior to Java, the main reason is that compared with Java, Python has simpler grammatical rules and language structures. The programs written by Python are closer to natural language queries. Therefore, the semantic mappings between the natural language queries and programs are easier for Python than Java.}

\begin{tcolorbox}[breakable,width=\linewidth,
boxrule=0pt,top=1pt, bottom=1pt, left=1pt,right=1pt, colback=gray!20,colframe=gray!20]
\textbf{\ding{45} $\blacktriangleright$Answer to RQ1$\blacktriangleleft$} The performance of the sequential-based approaches are inconsistent on Java and Python dataset, we contribute it to the missed structural information learnt by these networks. In contrast, \tool outperforms them significantly. Furthermore, the bidirectional message passing and multi-head attention module could improve the model learning capacity to produce better results as compared to other GNN variants.
\end{tcolorbox}

\subsection{RQ2: Ablation Study on Each Component in \tool.}\label{sec:ablation}

We conduct the ablation study to investigate the impact of each component \ie, BiGGNN and multi-head attention on the separate encoder to confirm the local structural information and the global dependencies in the graph are both beneficial for code search.

The experimental results are shown in Table~\ref{ablation-separate}, where the checkmark denotes the used component. We can observe that BiGGNN improves the performance significantly as compared to multi-head attention for the program encoder or the summary encoder. It is an interesting finding because compared with the graphs used in the program scenario~\cite{liu2020retrieval, allamanis2017learning, allamanis2020typilus, zhou2019devign}, the transformer architecture~\cite{vaswani2017attention}, where multi-head attention is the key component in it, is dominating the NLP community. Hence, encoding the code into graphs is a common practice for a program, however, it is not very common for the natural language, even there are some existing works~\cite{chen2019reinforcement, su2020multi, chen2019graphflow, song2018exploring, de2018question}. One possible reason we conjecture is that the standard transformer~\cite{vaswani2017attention} is equipped with the encoder-decoder architecture, which is more powerful for the text generation tasks such as machine translation~\cite{bahdanau2014neural, vaswani2017attention, luong2015effective}, text summarization~\cite{maybury1999advances, liu2019text}. The experimental results confirm the necessity to explore the structural information behind the query text for code search. Furthermore, we also find that the performance on Python dataset is still higher than Java when turning off some components, which indicates that the model is easier to learn semantics on Python rather than Java. \revised{We also calculate the p-value of \tool and its different components on  Java or Python testset respectively. The p-values are presented in the rightmost column of Table~\ref{ablation-separate}. We can observe that the majority of p-values are less than 0.1, which further confirms the effectiveness of each component in \tool.}

\revised{We further conduct an experiment to validate the effectiveness of our proposed encoder by replacing it with BiLSTM or transformer encoder for program or summary, respectively. Specifically, for BiLSTM encoder, we utilize the hidden states produced by a two-layer BiLSTM for the program or summary encoding, for the transformer encoder (abbr. Trans in Table~\ref{ablation-separate-1}), following Vaswani et al.~\cite{vaswani2017attention}, we utilize 6 encoder layers with 8 heads in each layer to encode the sequence and further utilize the max-pooling operation over the output to obtain the vector representations. The other settings are the same with \tool and experimental results are presented in Table~\ref{ablation-separate-1}. We can observe that when modeling programs using the graph encoder while modeling the summaries using BiLSTM or transformer, MRR is 63.64, 64.38 on Java dataset, 72.12, 72.88 on Python dataset which is lower than \tool, which demonstrates that the structural information in the query is beneficial for code search. Furthermore, we can also find that the values produced by replacing the program encoder with BiLSTM or transformer are also lower than \tool. Hence, to sum up, the results confirm that our proposed encoder for both programs and summaries can help the model achieve better performance.}


\begin{tcolorbox}[breakable,width=\linewidth,
boxrule=0pt,top=1pt, bottom=1pt, left=1pt,right=1pt, colback=gray!20,colframe=gray!20]
\textbf{\ding{45} $\blacktriangleright$Answer to RQ2$\blacktriangleleft$} BiGGNN and multi-head attention are both effective in improving the performance, however, the local structural information captured by BiGGNN is more critical. When incorporating both, \tool can get the best performance. 
\end{tcolorbox}

\begin{table*}[]
\centering
\caption{Ablation study of the performance on the encoder with different components on Java and Python data set.}
\label{ablation-separate}
\scriptsize {\addtolength{\tabcolsep}{2pt}
\begin{tabular}{c|cc|cc|ccccc|c}
\hline
\multirow{2}{*}{Dataset} & \multicolumn{2}{c|}{Program} & \multicolumn{2}{c|}{Summary} & \multirow{2}{*}{R@1} &\multirow{2}{*}{R@5}& \multirow{2}{*}{R@10} & \multirow{2}{*}{NDCG} & \multirow{2}{*}{MRR} & \multirow{2}{*}{\revised{p-value}} \\
\cline{2-5}
                          & Multi-Head      & BiGGNN       & Multi-Head        & BiGGNN        &    &&&         &                \\\hline
\multirow{5}{*}{Java}    &  \checkmark     &           &  \checkmark       & \checkmark       & 29.95 &55.18   &64.42  &46.59  & 41.74   & \revised{0.01}        \\
                         &                 &  \checkmark  &    \checkmark    &   \checkmark   & 45.51 &67.55&74.16&59.78     &    55.65    & \revised{0.09}      \\
                         &   \checkmark   &     \checkmark      &     \checkmark   &             & 25.00 &48.66&58.16&    40.91      &   36.19 & \revised{0.00} \\
                         &     \checkmark   &     \checkmark &         &    \checkmark         & 48.26 &70.75&77.15&62.77        &   58.51    & \revised{0.18}       \\
                         &     \checkmark   &     \checkmark &   \checkmark        &    \checkmark  & \textbf{56.99}&\textbf{76.03}&\textbf{81.15}&\textbf{69.47} &  \textbf{65.80} & -  \\ \hline
\multirow{5}{*}{Python}  &  \checkmark     &           &  \checkmark       & \checkmark       & 13.76 &35.27&46.63&28.53   &     24.52   &\revised{0.00}         \\
                         &                 &  \checkmark  &    \checkmark    &   \checkmark   & 55.55 &77.11&83.24&  69.28   &    65.41     &\revised{0.12}        \\
                     &   \checkmark   &\checkmark    &\checkmark   &     & 27.55 &52.49&61.89&43.89  &  39.23  &\revised{0.00}    \\
                         &     \checkmark   &     \checkmark &       &\checkmark  &60.03&80.90&  86.37&73.31  &  69.59    &\revised{0.27}  \\
                         &     \checkmark   &   \checkmark &   \checkmark  &  \checkmark  &\textbf{65.31} &\textbf{84.21}&\textbf{89.05}&\textbf{77.30} & \textbf{73.90} & -\\ \hline
\end{tabular}
}
\end{table*}

\begin{table*}[]
\centering
\caption{\revised{Ablation study of the performance when the encoder is replaced with BiLSTM and Transformer encoder on Java and Python data set.}}
\label{ablation-separate-1}
\scriptsize {\addtolength{\tabcolsep}{-2pt}
\begin{tabular}{c|ccc|ccc|ccccc}
\hline
\multirow{2}{*}{Dataset} & \multicolumn{3}{c|}{Program} & \multicolumn{3}{c|}{Summary} & \multirow{2}{*}{R@1} &\multirow{2}{*}{R@5}& \multirow{2}{*}{R@10} & \multirow{2}{*}{NDCG} & \multirow{2}{*}{MRR} \\
\cline{2-7}
                         &BiLSTM & Trans &  \tool      &BiLSTM & Trans & \tool        &    &&&         &                \\\hline
\multirow{5}{*}{Java}   & \revised{\checkmark}  &&               &&&  \revised{\checkmark}  & \revised{50.12} &\revised{70.66} & \revised{76.79} & \revised{63.64} & \revised{59.73}       \\ 
                        &&\revised{\checkmark} &        &&&  \revised{\checkmark} &     \revised{52.39}    & \revised{73.49}  & \revised{79.51}  & \revised{66.10} & \revised{62.04} \\  
                        &&     &   \revised{\checkmark}  &  \revised{\checkmark}  &&   & \revised{54.89} & \revised{73.79}  & \revised{79.17} & \revised{67.27} &\revised{63.64} \\  
                        &&       &   \revised{\checkmark}   &&\revised{\checkmark}&     &\revised{55.45} & \revised{74.88} & \revised{80.36}  & \revised{67.83} &\revised{64.38}          \\  
                        && &     \checkmark   &      &           &    \checkmark  & \textbf{56.99}&\textbf{76.03}&\textbf{81.15}&\textbf{69.47} &  \textbf{65.80}  \\ \hline
\multirow{5}{*}{Python}& \revised{\checkmark}  &&      &&         & \revised{\checkmark}& \revised{59.56} & \revised{80.28}&\revised{85.85}  &  \revised{72.69}&\revised{68.96}\\ 
                        &&\revised{\checkmark} &      &&   & \revised{\checkmark}   & \revised{61.63} & \revised{81.06} & \revised{86.25} &  \revised{74.55} & \revised{71.47}     \\  
                        &       &   &\revised{\checkmark}&\revised{\checkmark}& &    &  \revised{62.93}  & \revised{82.37} & \revised{87.87}  & \revised{75.02} & \revised{72.12}     \\  
                        &      &          & \revised{\checkmark}&& \revised{\checkmark}&    & \revised{63.46}  & \revised{82.58} &\revised{88.33}&  \revised{76.20} & \revised{72.88}\\ 
                        && &     \checkmark   &   &    &  \checkmark  &\textbf{65.31} &\textbf{84.21}&\textbf{89.05}&\textbf{77.30} & \textbf{73.90} \\ \hline
\end{tabular}
}
\end{table*}

\subsection{RQ3: Hop\&Head Analysis.}
\begin{figure}
     \centering
     \begin{subfigure}[b]{0.24\textwidth}
         \centering
         \includegraphics[width=\textwidth, height=3.7cm]{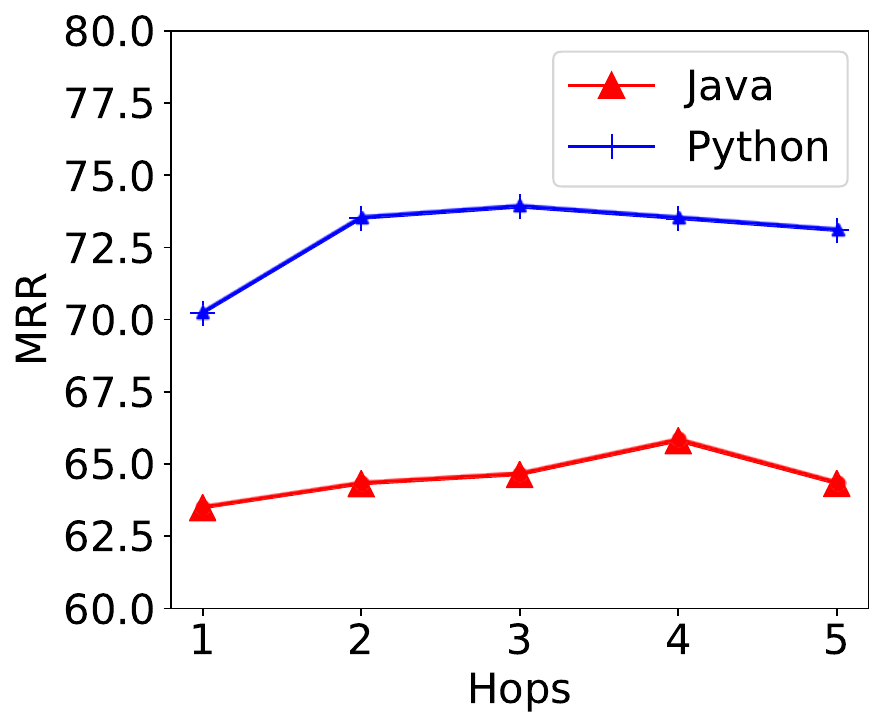}
         \caption{MRR of Different Hops}
          \label{fig:hop_mrr}
     \end{subfigure}
     \begin{subfigure}[b]{0.24\textwidth}
         \centering
         \includegraphics[width=\textwidth]{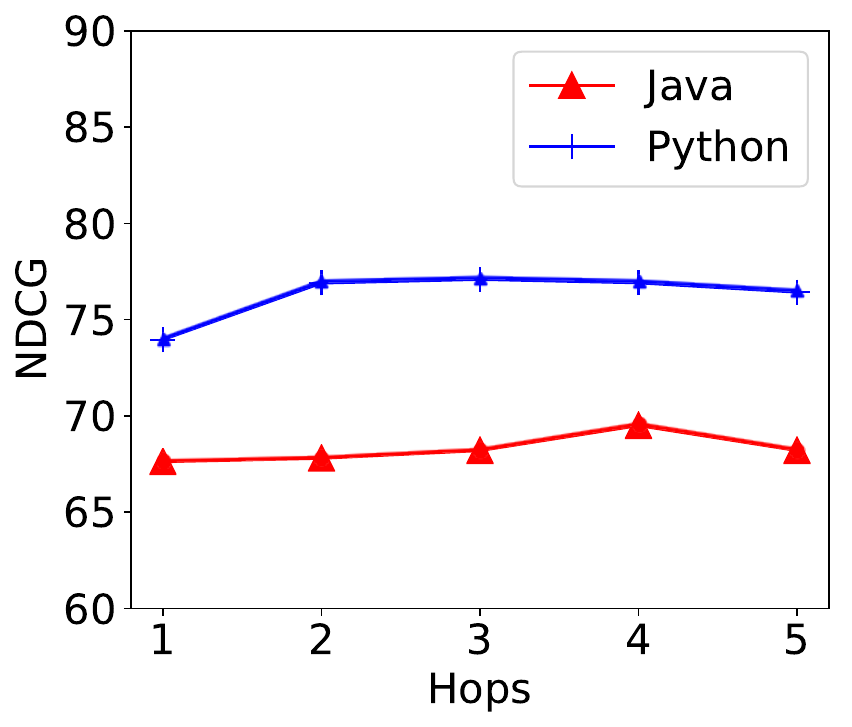}
         \caption{NDCG of Different Hops}
         \label{fig:hop_ndcg}
     \end{subfigure}
     \begin{subfigure}[b]{0.24\textwidth}
         \centering
         \includegraphics[width=\textwidth, height=3.7cm]{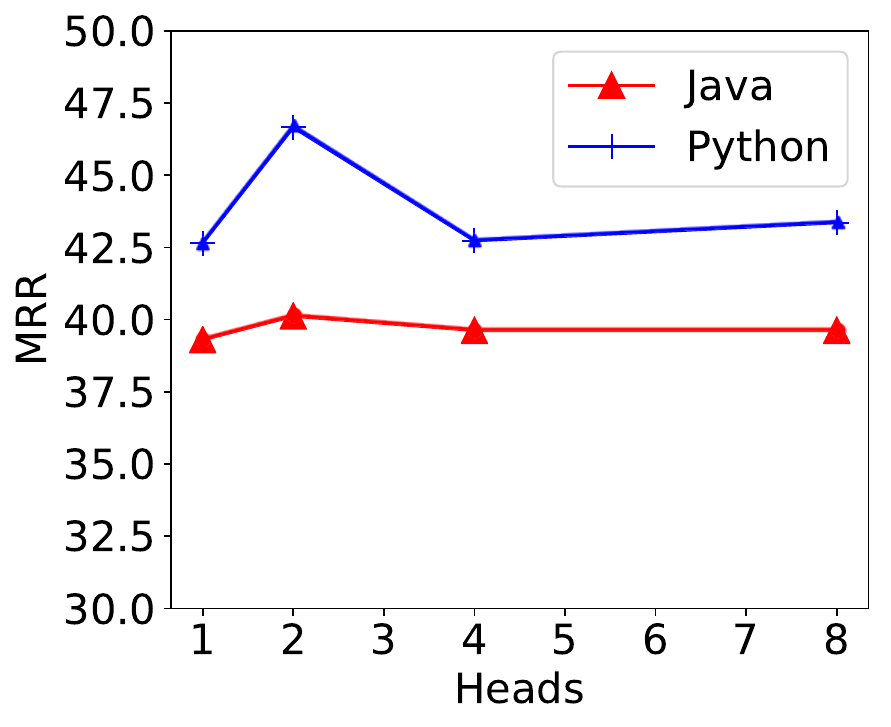}
         \caption{MRR of Different Heads}
         \label{fig:head_mrr}
     \end{subfigure}
     \begin{subfigure}[b]{0.24\textwidth}
         \centering
         \includegraphics[width=\textwidth]{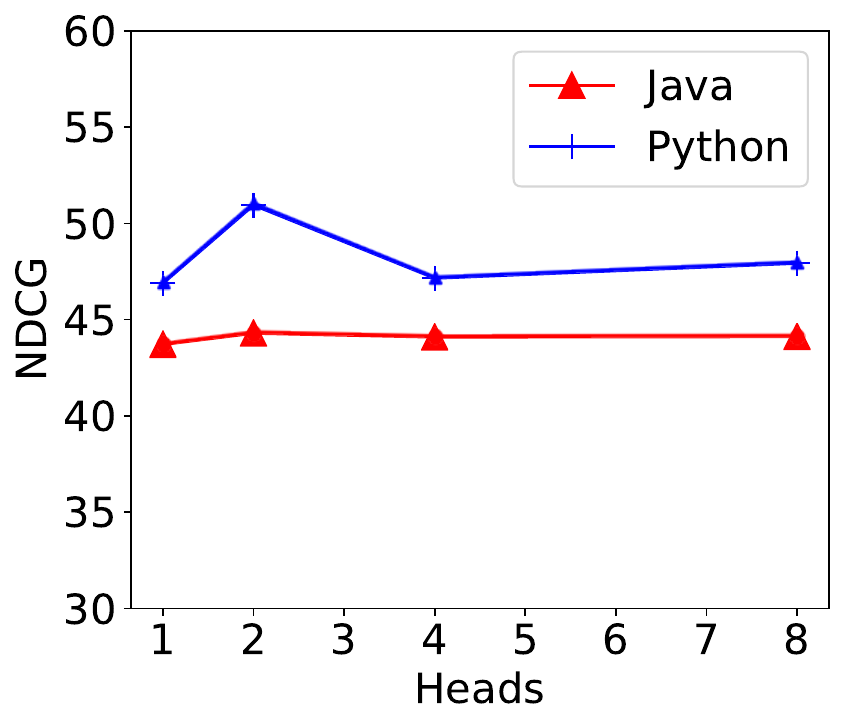}
         \caption{NDCG of Different Heads}
         \label{fig:head_ndcg}
     \end{subfigure}
        \caption{The effect of the number of hops and heads to capture the local structural information and the global dependencies. }
    \label{fig:hop}
\end{figure}

We further investigate the impact of the different number of hops (see $K$ in Eq.~\ref{eq:maxpool}) and heads (see $h$ in Eq.~\ref{eq:muti-head}) on the capacity to capture the local structural information and the global dependencies in the graph. Specifically, we set the number of hops in a range of 1 to 5 and heads in a range of 1 to 8 and keep the other settings unchanged for a fair comparison. Note that we only turn on BiGGNN with the multi-head attention closed for two encoders to analyse the impact of hops (and vice versa for head analysis), which is different to the settings in Section~\ref{sec:ablation}, where the specific module is closed for one encoder. Since the trend on R@1, R@5 and R@10 is similar to MRR and NDCG, we only show the results of MRR and NDCG in Fig.~\ref{fig:hop}.

We can see that in Fig.~\ref{fig:hop_mrr} and Fig.~\ref{fig:hop_ndcg}, with the increasing number of hops, MRR and NDCG improve gradually and they reach the highest at 4 and 3 for Java and Python respectively, after that the values begin to decrease, which is consistent with the findings that GNNs suffer from the over-squashing~\cite{alon2021on} when increasing the hops to learn the information from long-distant nodes. Hence, in our experiment, we set the value to 4 and 3 for Java and Python to achieve the best performance. The reason why Java needs more hops than Python is that the constructed graph for Java is larger than Python and the statistical distributions of nodes and edges on the dataset are shown in Table~\ref{tbl:graph_statistics}. We can observe that the average node size in the program graph and summary graph on the Java dataset is 125.99 and 17.44, which is larger than the Python dataset \ie, 100.35 and 14.00 respectively. Furthermore, the average connections (edges) of Python on the program graph is larger than Java \ie, 171.55 VS 125.22. Hence, the constructed graph for Python dataset tends to be smaller and better-connected, which makes the model require fewer hops than Java to achieve the best performance. For \revised{multi-head} attention to capture the global dependencies, from Fig.~\ref{fig:head_mrr} and Fig.~\ref{fig:head_ndcg}, we find that the trend is basically the same on Java and Python. When increasing the number of heads to 2, the values reach the highest, which indicates 2 heads are enough to learn the global dependencies in a graph. 

\begin{tcolorbox}[breakable,width=\linewidth,
boxrule=0pt,top=1pt, bottom=1pt, left=1pt,right=1pt, colback=gray!20,colframe=gray!20]
\textbf{\ding{45} $\blacktriangleright$Answer to RQ3$\blacktriangleleft$} The optimal value of hops relies on the graph size in the dataset. In \tool, we set the hops to 4 and 3 for Java and Python to achieve the best performance on the used dataset. The setting of the number of heads is consistent on both Java and Python datasets where 2 heads are enough to capture the global dependencies.
\end{tcolorbox}

\begin{table}[]
\centering
\caption{The statistics of the graph size in the constructed program graph and the summary graph for Java and Python dataset.}
\label{tbl:graph_statistics}
\begin{tabular}{ccccc|ccc}
\hline
\multicolumn{2}{c|}{\multirow{2}{*}{Dataset}}        & \multicolumn{3}{c|}{Program Graph} & \multicolumn{3}{c}{Summary Graph} \\ \cline{3-8} 
\multicolumn{2}{c|}{}                                & min       & max      & avg         & min        & max        & avg         \\ \cline{1-8} 
\multirow{2}{*}{Java}   & \multicolumn{1}{c|}{Node} & 55        & 200      & \textbf{125.99}      & 3          & 200        & \textbf{17.44}       \\
                        & \multicolumn{1}{c|}{Edge} & 54        & 263      & 125.22      & 2          & 422        & \textbf{31.29}       \\ \hline
\multirow{2}{*}{Python} & \multicolumn{1}{c|}{Node} & 10        & 200      & 100.35      & 3          & 199        & 14.00       \\
                        & \multicolumn{1}{c|}{Edge} & 9         & 477      & \textbf{171.55}      & 2          & 408        & 24.21       \\ \hline
\end{tabular}
\end{table}

\begin{table*}[]
\centering
\caption{\revised{The average cosine similarity score of top-1 results over the 99 real queries.}}
\label{tbl-avg_score}
\begin{tabular}{c|cc|cc|cc|cc|cc|cc}
\hline
\multirow{2}{*}{Model} & \multicolumn{2}{c|}{NBoW} & \multicolumn{2}{c|}{biRNN} & \multicolumn{2}{c|}{SelfAtt Encoder} & \multicolumn{2}{c|}{1D-CNN} & \multicolumn{2}{c|}{UNIF} & \multicolumn{2}{c}{\tool} \\
                       & Java       & Python       & Java        & Python       & Java             & Python            & Java        & Python        & Java       & Python       & Java            & Python            \\ \hline
Avg                    &   \revised{ 0.83}   &   \revised{0.81}       &    \revised{0.80}    &    \revised{0.88}     &   \revised{ 0.99}          &     \revised{1.01}         &   \revised{0.71}     &   \revised{ 0.79}      &  \revised{ 0.84}    &  \revised{0.81}      & \revised{\textbf{1.25}}            & \revised{\textbf{1.25}}              \\ \hline
\end{tabular}
\end{table*}

\begin{table*}[]
\centering
\caption{\revised{Experimental results on 99 queries with their annotated programs.}}
\label{table-baselines-annotations}
\begin{tabular}{c|ccccc|ccccc}
\hline
\multicolumn{1}{c|}{\multirow{2}{*}{Model}} & \multicolumn{5}{c|}{Java}        & \multicolumn{5}{c}{Python}   \\ \cline{2-11} 
\multicolumn{1}{c|}{}                       & R@1 & R@5 & R@10 & NDCG & MRR & R@1 & R@5 & R@10 & NDCG & MRR \\ \hline
NBoW                         &  \revised{79.79}  &  \revised{93.93}   &  \revised{94.94}  & \revised{84.81}  &  \revised{86.50}   &  \revised{77.77}   & \revised{95.95}  & \revised{96.96}   & \revised{75.42}    & \revised{86.00}  \\
biRNN                        &  \revised{70.70}  & \revised{94.94}  & \revised{97.97}   & \revised{89.97}   &  \revised{81.10}  &  \revised{66.66}  &  \revised{90.90}   &  \revised{93.93}    &  \revised{88.25}  &  \revised{77.10}   \\
SelfAtt Encoder              &  \revised{67.67}    & \revised{92.92}   &  \revised{94.94}  &  \revised{90.79}   &  \revised{78.30}   & \revised{75.75}   &  \revised{92.92}  &  \revised{93.93}   &\revised{87.46} &  \revised{83.40} \\
1D-CNN                       &  \revised{58.58}    &  \revised{90.90}   & \revised{97.97}  & \revised{84.90}  & \revised{72.90} &  \revised{52.52}  &   \revised{82.82}    &  \revised{88.88}   & \revised{78.29}   &  \revised{66.30}   \\
UNIF                         &  \revised{79.79}   &  \revised{96.96}&  \revised{96.96} & \revised{86.03} &  \revised{87.10}    &  \revised{78.78} &  \revised{94.94}   &  \revised{96.96} & \revised{77.08}  &\revised{86.20}\\ \hline
GraphSearchNet               &  \revised{\textbf{83.73}}   & \revised{\textbf{97.92}} &  \revised{\textbf{97.97}}   & \revised{\textbf{91.57}}& \revised{\textbf{88.33}} &  \revised{\textbf{78.88}}   & \revised{\textbf{96.91}}  &  \revised{\textbf{96.97}}    &  \revised{\textbf{89.73}}  & \revised{\textbf{87.29}}  \\ \hline
\end{tabular}
\end{table*}

\subsection{RQ4: Quantitative Analysis for Real 99 Queries.}

\revised{We compare GraphSearchNet with CodeSearchNet, which consists of NBoW, biRNN, SelfAtt Encoder and 1D-CNN, and UNIF to evaluate the performance on the real 99 queries because the selected baselines tend to have a better performance than others in Table~\ref{table-baselines}. Specifically, we return the top-1 result for each query by different approaches. We manually check these top-1 results by different approaches and find that they are related to the query. To measure the degree of correlation, we calculate the cosine similarity between the query vector produced by the summary encoder $f_s$ and the returned top-1 program vector produced by the program encoder $f_c$, and further average them over 99 queries to get the mean score. The mean similarity scores for these approaches are shown in Table~\ref{tbl-avg_score}. We can find that the cosine similarity scores produced by GraphSearchNet are higher than others, which indicates that the top-1 results returned by \tool are more relevant to the queries than baselines. We attribute it to the effectiveness of our approach to learning semantic relations between programs and queries.} 

\revised{CodeSearchNet~\cite{husain2019codesearchnet} further provides 99 queries with manually annotated programs for evaluation. We also compare \tool and the baselines on this evaluation set. Specifically, since one query in the total of 99 queries may have multiple annotated programs with different relevance scores in this evaluation set. Hence, for Java and Python, we first extract the annotated program with the highest relevance score for each query. If multiple programs share the same highest relevance score, we randomly select one program. In this way, we can construct a testset where each sample is a pair of query and its most relevant annotated program. We utilize this testset for evaluation. Since there are only 99 samples in this testset, hence the answer of each query is retrieved from 99 candidate programs instead of 1000 programs, which is used for the other experiments in the rest of the paper. We directly test the performance of different models on this testset and the experimental results are presented in Table~\ref{table-baselines-annotations}. We find that \tool still achieves better performance on this testset compared with baselines, which indicates the effectiveness of our approach on different testsets. In addition, we can observe that these values are improved greatly compared with the values in Table~\ref{table-baselines}, it is reasonable since the answer of each query is retrieved from 99 candidate programs, which is much easier for the model to distinguish than 1000 programs used in Table~\ref{table-baselines}.}

Furthermore, we show two examples of the returned top-1 results for Java and Python from the search codebase in Table~\ref{tbl-example-java} and Table~\ref{tbl-example-python} by \tool and the best baseline \ie, SelfAtt Encoder in Table~\ref{tbl-avg_score}. Both two examples, \tool can produce better retrieved results than baselines. The completed top-10 results of 99 queries are provided in our official repository. We find that given the query ``get executable path'' and ``how to determine a string is a valid word'', the generated results by SelfAtt Encoder are less relevant to the query on both Java and Python. In particular, for the query, ``get executable path'', the results by \tool are more accurate than SelfAtt Encoder to produce a program with the functionality of returning a path. For the query, ``how to determine a string is a valid word'', it aims at verifying whether a string is a word, the produced results by SelfAtt Encoder indicate that it misunderstood the query semantics and return the programs related to the password verification for both Java and Python. In contrast, \tool can produce more accurate and semantic-relevant programs based on the query. 
\revised{We also provide one example, which GraphSearchNet cannot provide the correct code snippet on Java and Python dataset. As shown in Table~\ref{tbl-example-failed}, for the query ``aes encryption'', the returned results for SelfAtt Encoder are more accurate than GraphSearchNet. Specifically, SelfAtt provides the accurate code snippet for the query, while \tool provides a decryption code snippet on Java dataset and provides an irrelevant function on Python dataset. The main reason is that \tool leverages dependency parsing to construct a graph to capture token relations in a query. However, for the query of `aes encryption'', since it is short and only has two tokens, the constructed graph is very simple without sufficient edges to fully express the relations between tokens. \tool cannot learn sufficient semantics and thus provide inaccurate results. We further conduct statistical analysis on the queries that \tool fails to provide accurate results over 99 queries. We find that in a total of 99 queries, 57 queries start with a verb such as ``convert int to string'', 12 queries start with the word how such as ``how to get current date'' and 11 queries start with the noun phrase such as ``aes encryption''. By comparing these queried results, we find that generally, \tool has relatively poor performance for noun phrases on Java and Python dataset compared with two other kinds of queries. The main reason is that the noun phrases are short. \tool cannot learn sufficient semantics when the provided query has limited information.}

\begin{tcolorbox}[breakable,width=\linewidth,
boxrule=0pt,top=1pt, bottom=1pt, left=1pt,right=1pt, colback=gray!20,colframe=gray!20]
\textbf{\ding{45} $\blacktriangleright$Answer to RQ4$\blacktriangleleft$} By calculating the average cosine similarity score over the returned top-1 program vectors with the real query vector, we find that \tool could produce more accurate and semantic-relevant programs. In addition, we present some examples as compared to the baselines for analysis. 
\end{tcolorbox}

\begin{table*}[t]
\centering
\caption{Two examples of the queried top-1 results for Java from the search codebase.}
\label{tbl-example-java}
\scriptsize {\addtolength{\tabcolsep}{2pt}
\begin{tabular}{c|l|l}
\hline
Query            &  \multicolumn{1}{c|}{SelfAtt Encoder} & \multicolumn{1}{c}{\tool} \\ \hline
    
get executable path
&  \begin{lstlisting} 
private Path getCollectionPath() {
        Path collectionPath = PathUtils.
        ReportsDirectory
        .getPathForObject(getInstance());
        return collectionPath.
        resolve(provider.
        getArtifactPath(getInstance()));
    }
\end{lstlisting}
& \begin{lstlisting} 
   public static String getPathToExecutable
   (File featureDir, String path) {
        File scriptPath = new File(path);
        if ( ! scriptPath.isAbsolute() ) {
            scriptPath = 
            new File(featureDir, path);
        }
        return scriptPath.getAbsolutePath();
    }
\end{lstlisting}     \\ \hline
how to determine a string is a valid word 
&  \begin{lstlisting} 
private static String promptForPassword
        (String passwordName, 
        String commandLineOption, String envVarName) 
        throws Exception {
        final Console console = System.console();
        if(console == null) {
            throw new Exception("Cannot allocate 
            a console. Set env var "+envVarName+" 
            or "+commandLineOption+" 
              on commandline in that case");
        }
        return new String(console.
        readPassword("[%s]", passwordName+
          " password:"));
    }
\end{lstlisting}
& \begin{lstlisting} 
    public static boolean isWord 
         (@Nullable final String sStr) {
        if (StringHelper.hasNoText (sStr))
            return false;
        return isWord (sStr.toCharArray ());
    }
\end{lstlisting}     \\ \hline
\end{tabular}
}
\end{table*}

\begin{table*}[t]
\centering
\caption{Two examples of the queried top-1 results for Python from the search codebase.}
\label{tbl-example-python}
\scriptsize {\addtolength{\tabcolsep}{2pt}
\begin{tabular}{c|l|l}
\hline
Query            &  \multicolumn{1}{c|}{SelfAtt Encoder} & \multicolumn{1}{c}{\tool} \\ \hline
    
get executable path
&  \begin{lstlisting} 
def get_mp_bin_path():
   plat = platform.system()
   if plat == "Linux":
     return resource_filename(__name__, 
     "bin/metaparticle/linux/mp-compiler")
   elif plat == "Windows":
     return resource_filename(__name__, 
     "bin/metaparticle/windows/mp-compiler.exe")
   elif plat == "Darwin":
     return resource_filename(__name__, 
     "bin/metaparticle/darwin/mp-compiler")
   else:
     raise Exception("Your platform 
           is not supported.")
\end{lstlisting}
& \begin{lstlisting} 
   def get_path():
        return os.path.abspath(os.path.
        dirname(os.path.dirname(__file__)))
\end{lstlisting}     \\ \hline
how to determine a string is a valid word 
&  \begin{lstlisting} 
def _validate_admin_password(admin_password):
    password_regex = r"[A-Za-z0-9@#$%^&+=]{6,}"
    pattern = re.compile(password_regex)
    if not pattern.match(admin_password):
        raise ValueError(red(
            "The password must be at 
            least 6 characters and 
            contain only the 
            following characters:\n"
            "A-Za-z0-9@#$%^&+="
        ))
    return admin_password
\end{lstlisting}
& \begin{lstlisting} 
    def isValid(self, text, word):
      return bool(re.search(word, text, 
          re.IGNORECASE))
\end{lstlisting}     \\ \hline
\end{tabular}
}
\end{table*}

\begin{table*}[]
\centering
\caption{\revised{One example of results for the query ``aes encryption'' from the search codebase.}}
\label{tbl-example-failed}
\scriptsize {\addtolength{\tabcolsep}{2pt}
\begin{tabular}{c|l|l}
\hline
Dataset            &  \multicolumn{1}{c|}{SelfAtt Encoder} & \multicolumn{1}{c}{GraphSearchNet} \\ \hline
Java
&  \begin{lstlisting} 
public static byte[] aesEncrypt(byte[] input, 
    byte[] key, byte[] iv) throws Exception {
		return aesEncrypt(input, new SecretKeySpec
		   (key, "AES"), iv);
	}
\end{lstlisting} 
& \begin{lstlisting} 
  public static String aesDecrypt(byte[] input,
    byte[] key) {
		byte[] decryptResult = aes(input, key, 
		    Cipher.DECRYPT_MODE);
		return new 
		   String(decryptResult, Charsets.UTF_8);
	}
\end{lstlisting}  \\ \hline
Python
&  \begin{lstlisting} 
def _aes_encrypt(self, text, key):
    pad = 16 - len(text) % 16
    text = text + pad * chr(pad)
    encryptor = AES.new(key, 2, 
         '0102030405060708')
    enc_text = encryptor.encrypt(text)
    enc_text_encode = base64.b64encode(enc_text)
    return enc_text_encode
\end{lstlisting}
& \begin{lstlisting} 
  def _maybe_add_hash(tsig_alg, hash_alg):
    try:
       _hashes[tsig_alg] = dns.hash.get(hash_alg)
    except KeyError:
       pass
\end{lstlisting}  \\ \hline
\end{tabular}
}
\end{table*}
\section{Discussion}\label{sec:discussion}
This section presents the impact of the dimensional size used in \tool. We further discuss the widely used pre-trained models in the program scenario and followed by the threats to the validity of our work.  
\subsection{Impact of Dimensional Size}
\begin{figure}
     \centering
     \begin{subfigure}[b]{0.24\textwidth}
         \centering
         \includegraphics[width=\textwidth]{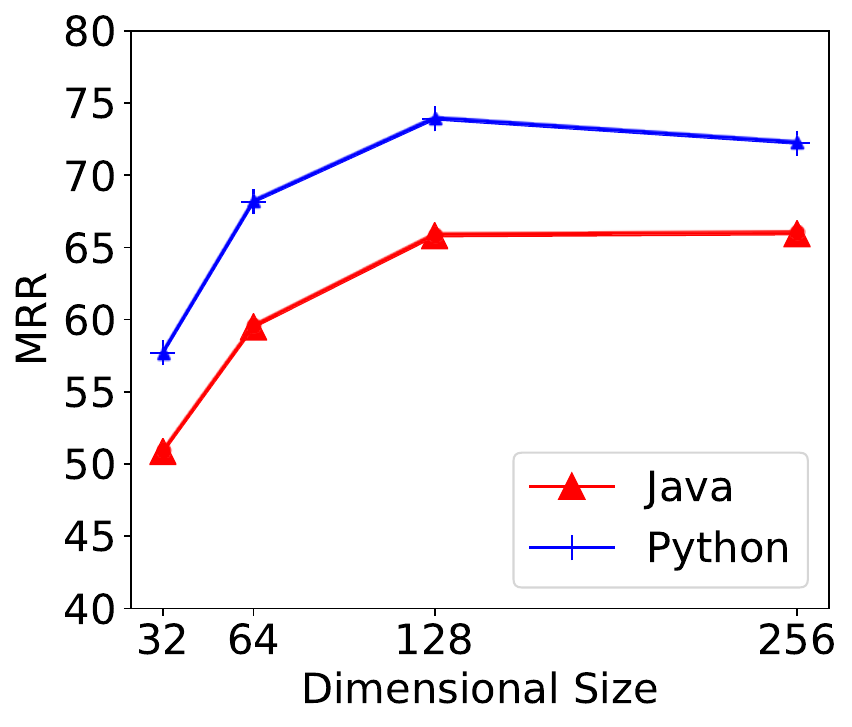}
         \caption{MRR}
          \label{fig:dim_mrr}
     \end{subfigure}
     \begin{subfigure}[b]{0.24\textwidth}
         \centering
         \includegraphics[width=\textwidth]{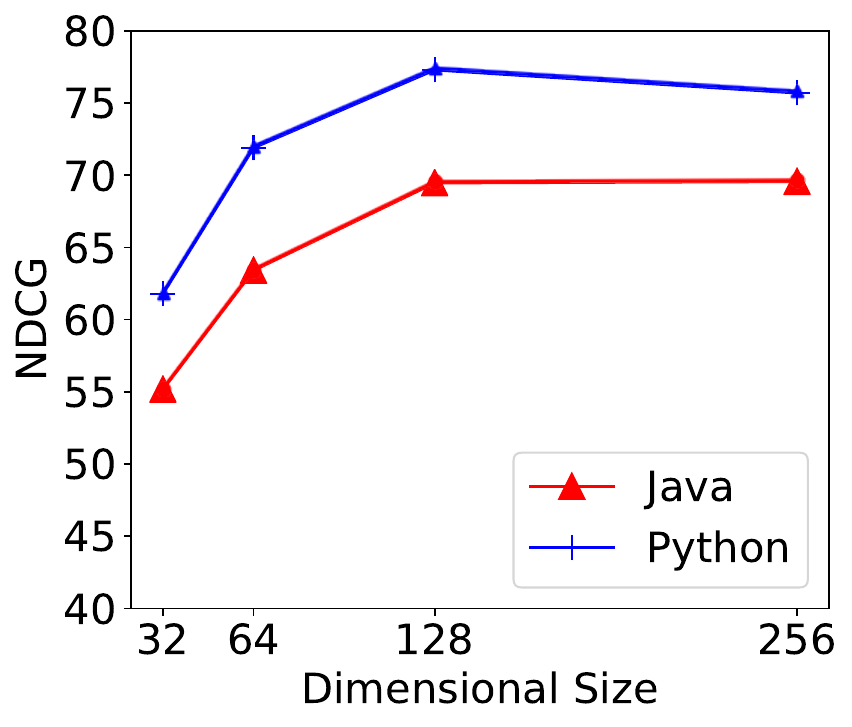}
         \caption{NDCG}
         \label{fig:dim_ndcg}
     \end{subfigure}
        \caption{The effect of the dimensional size with regard to MRR and NDCG.}
    \label{fig:dim}
\end{figure}

We study the impact of different dimensional sizes on the performance of \tool. We only change the dimensional size (32/64/128/256) and keep the remaining settings unchanged for the evaluation. The results are shown in Fig.~\ref{fig:dim}. We can observe that the performance improves greatly as the feature size increases since the learning capacity of the model is increased with the growing dimension size. However, we can also see that after the 128-dimensional size, MRR and NDCG on Java dataset improve slightly, while on Python dataset, there is an obvious decrease from the highest. We conjecture this is caused by when set to a higher dimensional size, the learning capacity is further increased to make the model overfit to the Python dataset, which harms the performance on Python dataset. In addition, with the increasing dimensional size, the training time and GPU memory consumption are also increased. For unification, we set the dimensional size to 128 on both Java and Python datasets for the evaluation.

\subsection{Pre-trained Models}
\revised{Recently, there are some unsupervised pre-trained approaches e.g., CodeBERT~\cite{feng2020codebert} and GraphCodeBERT~\cite{guo2020graphcodebert} that aim at learning
the general program representations for a variety of code-related tasks. Code search is selected as one of the downstream tasks. We also compare the performance of \tool with CodeBERT and GraphCodeBERT. Specifically, we use the officially released code by CodeBERT~\cite{feng2020codebert} and GraphCodeBERT~\cite{guo2020graphcodebert} with the default hyper-parameters to validate the performance on our dataset. Similar to CodeSearchNet~\cite{husain2019codesearchnet} and \tool, we set the batch size on the testset equal to 1000, which means the answer of each query is retrieved from 1000 candidate programs. Following CodeBERT and GraphCodeBERT, we use MRR as the evaluation metric and the experimental results are presented in Table~\ref{table-codebert}. We can observe that the MRR values of \tool are higher than CodeBERT, which demonstrates that utilizing structural information in code and query can help the model achieve better performance than CodeBERT, which only treats them as the sequences, even \tool has limited data for training. However, the MRR values of \tool are slightly lower than GraphCodeBERT. On one hand, it demonstrates that using structural information (such as dataflow in GraphCodeBERT) is effective. On other hand, GraphCodeBERT utilizes 2.3M functions for pre-training and the model has a total number of 125M parameters. In contrast, \tool only uses 0.25M data to train and the model only has 2M parameters. Considering the scale of the used data (0.25M VS 2.3M) and the model parameters (2M VS 125M), we believe \tool achieves comparable performance with GraphCodeBERT.}

\begin{table}[!t]
\centering
\caption{\revised{The MRR values as compared to CodeBERT and GraphCodeBERT.}}
\label{table-codebert}
\begin{tabular}{c|ccc}
\hline
\revised{Dataset} & \revised{CodeBERT} & \revised{GraphCodeBERT} & \revised{GraphSearchNet} \\ \hline
\revised{Java}    &  \revised{64.44}      &    \textbf{\revised{66.21}}           &      \revised{65.80}         \\ \hline
\revised{Python}  &    \revised{73.47}      &    \textbf{\revised{74.87}}           &    \revised{73.90}           \\ \hline
\end{tabular}
\end{table}

\subsection{Threats to Validity}
The internal threats to validity lie in our implementations, the graph construction for the program and the summary, the model implementation. To reduce this threat, we utilize the open-source tool~\cite{features-javac,fernandes2018structured} for Java and Python program graph construction, Spacy~\cite{spacy2} for the summary graph construction. To reduce the implementation threat, the co-authors carefully check the correctness of our implementation. We further make implementation public at \url{https://github.com/shangqing-liu/GraphSearchNet} for further investigation. 

The external threats to validity include the selected datasets and the evaluation of the baselines and the evaluation metrics. To reduce the impact of the dataset, we select the Java and Python datasets from CodeSearchNet~\cite{husain2019codesearchnet}, which has been widely used for other works~\cite{feng2020codebert, guo2020graphcodebert, wang2021codet5, lu2021codexglue} to evaluate the model performance. For the evaluation of baselines, we select the open-sourced baselines~\cite{husain2019codesearchnet, cambronero2019deep, gu2018deep, shuai2020improving, xu2021two, yao2019coacor} and evaluate the performance based on the default hyper-parameters used in the original papers. We consider these default parameters are optimal. We carefully checked the released code to mitigate the threat and plan to evaluate more hyper-parameters in the future. For the evaluation metrics, there are some other metrics rather than what we used such as mean average precision (MAP)~\cite{schutze2008introduction}, which can be used to evaluate the retrieval system, however, we follow the existing works~\cite{husain2019codesearchnet, cambronero2019deep, gu2018deep, shuai2020improving, xu2021two, yao2019coacor} in code search and select the widely used SuccessRate@k, MRR and NDCG for the evaluation.

\section{Related Work}\label{sec:related}
This section summarizes the related work on code search, graph neural networks, and program representation techniques.
\subsection{Code Search} 
Early works~\cite{lu2015query,lv2015codehow, bajracharya2006sourcerer, bajracharya2012analyzing, krugler2013krugle, liu2021codematcher} in code search utilized information retrieval ({IR}) to query the code by extracting both query and code characteristics. For example, Lu et al. ~\cite{lu2015query} extended a query with some synonyms generated from WordNet to improve the hit ratio. McMillan et al.~\cite{mcmillan2011portfolio} proposed Portfolio, a code search engine that combines keyword matching to return functions. CodeHow~\cite{lv2015codehow} extended the query with some APIs and utilized them with a boolean model to retrieve the matched programs. Despite these works could produce some accurate results, the performance of proposed approaches mainly relies on the searched code base and the \revised{semantic mapping} between the program and query is not addressed.

To learn semantics for the program and the query, some deep learning-based approaches were proposed~\cite{gu2018deep,husain2019codesearchnet,cambronero2019deep, shuai2020improving, haldar-etal-2020-multi, yao2019coacor, fang2021self, ishtiaq2021bert2code, du2021single} and these works attempted to use the deep learning techniques \eg, LSTMs, CNNs, Transformer to learn high-dimensional representations for programs and queries. For example, DeepCS~\cite{gu2016deep} encoded API sequence, tokens, and function names with multiple LSTMs to get the vector representations, and further encoded the queries into another vector with an extra LSTM for code search. CodeSearchNet~\cite{husain2019codesearchnet} explored some basic encoders such as LSTM, CNN, SelfAtt Encoder for code search. Furthermore, UNIF~\cite{cambronero2019deep} proposed that using a simple bag-of-word model with an attention mechanism can achieve state-of-the-art performance. However, these DL-based techniques in code search are mostly based on sequential models and ignore the rich structural information behind the programs and queries. In contrast, \tool attempts to extract structural information hidden in the text for the program and the query and further constructs the graphs to capture the semantic relations between them. \revised{OCoR~\cite{zhu2020ocor} proposed a neural network code search. Specifically, it represented each word by combing the vector representations of its characters to capture the overlap between the names. Furthermore, it designed an overlap matrix to represent the overlap degree between words in query and identifiers in code. Based on the character embedding and the overlap matrix, it further utilized a neural network for code search and utilized MRR to evaluate the model performance. Compared with this work, we design a new neural network architecture to fully utilize the structural information hidden in the program and query to learn semantics. Besides MRR, we further add R@1/5/10 and NDCG as the evaluation metrics.} Another work MMAN~\cite{wan2019multi} encoded the program sequence, AST and CFG on a manually collected C dataset with LSTM, Tree-LSTM and GGNN to learn the representation of a program. It further encoded the query with another LSTM for multi-modal learning. However, the relevant tools for graph construction are not available yet and we can not reproduce the experiments on our dataset. In addition, we encode the program and the query into graphs on Java and Python datasets and further improve the learning capacity of GNNs by BiGGNN and multi-head attention to achieve the best performance. 

\subsection{Graph Neural Networks}
GNNs~\cite{hamilton2017inductive,li2015gated, kipf2016semi, xu2018graph2seq} have attracted wide attention due to their power of learning structure data. Various applications from different domains such as chemistry biology~\cite{duvenaud2015convolutional}, computer vision~\cite{norcliffe2018learning}, natural language processing~\cite{xu2018graph2seq, chen2019reinforcement} have demonstrated the effectiveness of GNNs. GNNs have also been applied to code-related tasks. Compared with the early works to represent programs with abstract syntax tree~\cite{alon2018code2seq, alon2019code2vec, liu2020atom}, more works have shifted to use graphs~\cite{allamanis2017learning} to learn semantics for various tasks, such as source code summarization~\cite{liu2020retrieval, fernandes2018structured}, vulnerability detection~\cite{zhou2019devign}, type inference~\cite{allamanis2020typilus}, neural program repair~\cite{hellendoorn2019global}. For instance, Allamnanis et al.~\cite{allamanis2017learning} employed Gated Graph Neural Network (GGNN) in predicting the wrong usage of variables in programs. They further extended the similar approach to type inference~\cite{allamanis2020typilus}. Liu et al.~\cite{liu2020retrieval} proposed to combine GNNs with the retrieval information to enhance the generation of code summaries. Inspired by these state-of-the-art works that represent programs with graphs to capture the program semantics, in \tool, we propose to convert the programs and queries into graphs with BiGGNN to learn the structure information for semantic search. In addition, we also enhance BiGGNN by capturing the global dependencies for the nodes in a graph that is missed in the conventional GNNs. \revised{Hellendoorn et al.~\cite{hellendoorn2019global} proposed two hybrid model families (i.e., Graph Sandwiches and GREAT) to incorporate program structural and global information to learn programs for the task of variable-misuse localization and repair. Graph Sandwiches wrapped traditional gated graph message passing layers in sequential message passing layers and proposed the building block of [RNN, GGNN, RNN] to learn programs. GREAT generalized the relative position embeddings in Transformer by changing the attention computation to $e_{ij} = (\boldsymbol{q_i} + b_{ij})\boldsymbol{k_j}^T/\sqrt{N}$ to convey structural relations for learning. Compared with this work, we utilize bidirectional GGNN to learn program/query structures and further concatenate multi-head attention over any pair of nodes to capture the global dependencies in the graph for deep code search. The extensive experiments compared with the sequential models have demonstrated the effectiveness of \tool.}

\subsection{Program Representation}
Program representation, aiming at capturing the syntax and semantics behind the code, is a fundamental yet far-from-settled problem for code-intelligent tasks. The early works used feature selection and machine learning approaches such as n-gram model~\cite{frantzeskou2008examining},  SVM~\cite{linares2014using} for classifying source code. For instance, Linares et al.~\cite{linares2014using} utilized SVM and API information in classifying the software into their categories. Later, deep learning techniques were applied in this area to learn the general representations~\cite{mou2014convolutional, alon2019code2vec, allamanis2017learning}. Since abstract syntax tree (AST) is the high abstraction of programs, many works represented programs based on AST~\cite{mou2014convolutional, alon2019code2vec, alon2018code2seq, liu2020atom, astnn}. For example, TBCNN~\cite{mou2014convolutional} proposed a custom convolution neural network on AST to learn the program representation. CDLH~\cite{wei2017supervised} incorporated Tree-LSTM on AST for clone detection. Code2Vec~\cite{alon2019code2vec} and Code2Seq~\cite{alon2018code2seq} learnt the program representation with random sampled AST paths. In terms of graph-based techniques, Allamanis et al.~\cite{allamanis2017learning} innovated program representations by defining various types of edges on AST and utilized the Gated Graph Neural Network for learning program semantics and achieved state-of-the-art performance. Some follow-up works also adopted this paradigm and employed it on different tasks~\cite{liu2020retrieval, zhou2019devign}, such as source code summarization~\cite{fernandes2018structured,liu2020retrieval}, neural program repair~\cite{hellendoorn2019global}. Different from these works, which purely built the graph for programs, we also construct the query graph for code search. Furthermore, we also explore improving the learning capacity of conventional GNNs.

\section{Conclusion}\label{sec:conclusion}
In this paper, we propose \tool, a novel graph-based approach for code search to capture the semantic relations of the source code and the query. We construct the program graph and the summary graph with Bidirectional Graph Neural Network \ie, BiGGNN to learn the local structural information hidden in the sequential text. We further employ multi-head attention to capture the global dependencies that BiGGNN fails to learn in a graph. The extensive experiments on Java and Python datasets demonstrate the effectiveness of \tool. 
\section{ACKNOWLEDGMENTS}
This research is partially supported by the National Research Foundation, Singapore under its the AI Singapore Programme (AISG2-RP-2020-019), the National Research Foundation, Prime Ministers Office, Singapore under its National Cybersecurity R\&D Program (Award No. NRF2018NCR-NCR005-0001), NRF Investigatorship NRF-NRFI06-2020-0001, the National Research Foundation through its National Satellite of Excellence in Trustworthy Software Systems (NSOE-TSS) project under the National Cybersecurity R\&D (NCR) Grant award no. NRF2018NCR-NSOE003-0001, the Ministry of Education, Singapore under its Academic Research Tier 3 (MOET32020-0004). IIE authors are supported in part by NSFC (61902395), Beijing Nova Program and Alibaba Innovation Research. Any opinions, findings and conclusions or recommendations expressed in this material are those of the author(s) and do not reflect the views of the Ministry of Education, Singapore.
\bibliographystyle{IEEEtran}
\bibliography{ref.bib}
\end{document}